\documentclass[longauth]{aaEC}

\usepackage{graphicx}
\usepackage{natbib}
\usepackage{scalerel}
\usepackage{siunitx}
\usepackage{lastpage}
\usepackage{soul}

\usepackage[table]{xcolor}

\newcolumntype{P}[1]{>{\centering\arraybackslash}p{#1}}
\newcommand*{\rom}[1]{\expandafter\@slowromancap\romannumeral #1@}

\usepackage{txfonts}
\usepackage[pdfencoding=auto,psdextra]{hyperref}
\hypersetup{
    colorlinks=true,
    linkcolor=blue,
    filecolor=magenta,      
    urlcolor=blue,
    citecolor=blue
}
\makeatletter
\renewcommand*\aa@pageof{, page \thepage{} of \pageref*{LastPage}}
\makeatother

\usepackage[utf8]{inputenc}
\usepackage{datetime2}

\usepackage[switch, modulo]{lineno}
\linenumbers

\usepackage{euclid}

\begin{document}
\nolinenumbers
%
%
   \title{\Euclid: Searches for strong gravitational lenses using convolutional neural nets in Early Release Observations
   of the Perseus field
   \thanks{This paper is published on
     behalf of the Euclid Consortium.}}


\newcommand{\orcid}[1]{} 
\author{R.~Pearce-Casey\thanks{\email{rpc256@open.ac.uk}}\inst{\ref{aff1}}
\and B.~C.~Nagam\orcid{0000-0002-3724-7694}\inst{\ref{aff2}}
\and J.~Wilde\orcid{0000-0002-4460-7379}\inst{\ref{aff1}}
\and V.~Busillo\orcid{0009-0000-6049-1073}\inst{\ref{aff3},\ref{aff4},\ref{aff5}}
\and L.~Ulivi\orcid{0009-0001-3291-5382}\inst{\ref{aff6},\ref{aff7},\ref{aff8}}
\and I.~T.~Andika\orcid{0000-0001-6102-9526}\inst{\ref{aff9},\ref{aff10}}
\and A.~Manj\'on-Garc\'ia\orcid{0000-0002-7413-8825}\inst{\ref{aff11}}
\and L.~Leuzzi\orcid{0009-0006-4479-7017}\inst{\ref{aff12},\ref{aff13}}
\and P.~Matavulj\orcid{0000-0003-0229-7189}\inst{\ref{aff14}}
\and S.~Serjeant\orcid{0000-0002-0517-7943}\inst{\ref{aff1}}
\and M.~Walmsley\orcid{0000-0002-6408-4181}\inst{\ref{aff15},\ref{aff16}}
\and J.~A.~Acevedo~Barroso\orcid{0000-0002-9654-1711}\inst{\ref{aff17}}
\and C.~M.~O'Riordan\orcid{0000-0003-2227-1998}\inst{\ref{aff10}}
\and B.~Cl\'ement\orcid{0000-0002-7966-3661}\inst{\ref{aff17},\ref{aff18}}
\and C.~Tortora\orcid{0000-0001-7958-6531}\inst{\ref{aff3}}
\and T.~E.~Collett\orcid{0000-0001-5564-3140}\inst{\ref{aff19}}
\and F.~Courbin\orcid{0000-0003-0758-6510}\inst{\ref{aff20},\ref{aff21}}
\and R.~Gavazzi\orcid{0000-0002-5540-6935}\inst{\ref{aff22},\ref{aff23}}
\and R.~B.~Metcalf\orcid{0000-0003-3167-2574}\inst{\ref{aff12},\ref{aff13}}
\and R.~Cabanac\orcid{0000-0001-6679-2600}\inst{\ref{aff24}}
\and H.~M.~Courtois\orcid{0000-0003-0509-1776}\inst{\ref{aff25}}
\and J.~Crook-Mansour\orcid{0000-0001-7466-1192}\inst{\ref{aff26}}
\and L.~Delchambre\orcid{0000-0003-2559-408X}\inst{\ref{aff27}}
\and G.~Despali\orcid{0000-0001-6150-4112}\inst{\ref{aff12},\ref{aff13},\ref{aff28}}
\and L.~R.~Ecker\inst{\ref{aff29},\ref{aff30}}
\and A.~Franco\orcid{0000-0002-4761-366X}\inst{\ref{aff31},\ref{aff32},\ref{aff33}}
\and P.~Holloway\orcid{0009-0002-8896-6100}\inst{\ref{aff34}}
\and K.~Jahnke\orcid{0000-0003-3804-2137}\inst{\ref{aff35}}
\and G.~Mahler\orcid{0000-0003-3266-2001}\inst{\ref{aff27},\ref{aff36},\ref{aff37}}
\and L.~Marchetti\orcid{0000-0003-3948-7621}\inst{\ref{aff26},\ref{aff38},\ref{aff39}}
\and A.~Melo\orcid{0000-0002-6449-3970}\inst{\ref{aff10},\ref{aff9}}
\and M.~Meneghetti\orcid{0000-0003-1225-7084}\inst{\ref{aff13},\ref{aff28}}
\and O.~M\"uller\orcid{0000-0003-4552-9808}\inst{\ref{aff17}}
\and A.~A.~Nucita\inst{\ref{aff32},\ref{aff31},\ref{aff33}}
\and J.~Pearson\orcid{0000-0001-8555-8561}\inst{\ref{aff1}}
\and K.~Rojas\orcid{0000-0003-1391-6854}\inst{\ref{aff19}}
\and C.~Scarlata\orcid{0000-0002-9136-8876}\inst{\ref{aff40}}
\and S.~Schuldt\orcid{0000-0003-2497-6334}\inst{\ref{aff41},\ref{aff42}}
\and D.~Sluse\orcid{0000-0001-6116-2095}\inst{\ref{aff27}}
\and S.~H.~Suyu\orcid{0000-0001-5568-6052}\inst{\ref{aff9},\ref{aff10}}
\and M.~Vaccari\orcid{0000-0002-6748-0577}\inst{\ref{aff38},\ref{aff43},\ref{aff39}}
\and S.~Vegetti\orcid{0009-0006-0592-2882}\inst{\ref{aff10}}
\and A.~Verma\orcid{0000-0002-0730-0781}\inst{\ref{aff34}}
\and G.~Vernardos\orcid{0000-0001-8554-7248}\inst{\ref{aff44},\ref{aff45}}
\and M.~Bolzonella\orcid{0000-0003-3278-4607}\inst{\ref{aff13}}
\and M.~Kluge\orcid{0000-0002-9618-2552}\inst{\ref{aff30}}
\and T.~Saifollahi\orcid{0000-0002-9554-7660}\inst{\ref{aff46}}
\and M.~Schirmer\orcid{0000-0003-2568-9994}\inst{\ref{aff35}}
\and C.~Stone\orcid{0000-0002-9086-6398}\inst{\ref{aff47}}
\and A.~Paulino-Afonso\orcid{0000-0002-0943-0694}\inst{\ref{aff48},\ref{aff49}}
\and L.~Bazzanini\orcid{0000-0003-0727-0137}\inst{\ref{aff50},\ref{aff13}}
\and N.~B.~Hogg\orcid{0000-0001-9346-4477}\inst{\ref{aff51}}
\and L.~V.~E.~Koopmans\orcid{0000-0003-1840-0312}\inst{\ref{aff2}}
\and S.~Kruk\orcid{0000-0001-8010-8879}\inst{\ref{aff52}}
\and F.~Mannucci\orcid{0000-0002-4803-2381}\inst{\ref{aff8}}
\and J.~M.~Bromley\orcid{0000-0002-0304-096X}\inst{\ref{aff53}}
\and A.~D\'iaz-S\'anchez\orcid{0000-0003-0748-4768}\inst{\ref{aff11}}
\and H.~J.~Dickinson\orcid{0000-0003-0475-008X}\inst{\ref{aff1}}
\and D.~M.~Powell\orcid{0000-0002-4912-9943}\inst{\ref{aff10}}
\and H.~Bouy\orcid{0000-0002-7084-487X}\inst{\ref{aff54}}
\and R.~Laureijs\inst{\ref{aff55},\ref{aff2}}
\and B.~Altieri\orcid{0000-0003-3936-0284}\inst{\ref{aff52}}
\and A.~Amara\inst{\ref{aff56}}
\and S.~Andreon\orcid{0000-0002-2041-8784}\inst{\ref{aff57}}
\and C.~Baccigalupi\orcid{0000-0002-8211-1630}\inst{\ref{aff58},\ref{aff59},\ref{aff60},\ref{aff61}}
\and M.~Baldi\orcid{0000-0003-4145-1943}\inst{\ref{aff62},\ref{aff13},\ref{aff28}}
\and A.~Balestra\orcid{0000-0002-6967-261X}\inst{\ref{aff63}}
\and S.~Bardelli\orcid{0000-0002-8900-0298}\inst{\ref{aff13}}
\and P.~Battaglia\orcid{0000-0002-7337-5909}\inst{\ref{aff13}}
\and D.~Bonino\orcid{0000-0002-3336-9977}\inst{\ref{aff64}}
\and E.~Branchini\orcid{0000-0002-0808-6908}\inst{\ref{aff65},\ref{aff66},\ref{aff57}}
\and M.~Brescia\orcid{0000-0001-9506-5680}\inst{\ref{aff4},\ref{aff3},\ref{aff5}}
\and J.~Brinchmann\orcid{0000-0003-4359-8797}\inst{\ref{aff49},\ref{aff67}}
\and A.~Caillat\inst{\ref{aff22}}
\and S.~Camera\orcid{0000-0003-3399-3574}\inst{\ref{aff68},\ref{aff69},\ref{aff64}}
\and V.~Capobianco\orcid{0000-0002-3309-7692}\inst{\ref{aff64}}
\and C.~Carbone\orcid{0000-0003-0125-3563}\inst{\ref{aff42}}
\and J.~Carretero\orcid{0000-0002-3130-0204}\inst{\ref{aff70},\ref{aff71}}
\and S.~Casas\orcid{0000-0002-4751-5138}\inst{\ref{aff72},\ref{aff19}}
\and M.~Castellano\orcid{0000-0001-9875-8263}\inst{\ref{aff73}}
\and G.~Castignani\orcid{0000-0001-6831-0687}\inst{\ref{aff13}}
\and S.~Cavuoti\orcid{0000-0002-3787-4196}\inst{\ref{aff3},\ref{aff5}}
\and A.~Cimatti\inst{\ref{aff74}}
\and C.~Colodro-Conde\inst{\ref{aff75}}
\and G.~Congedo\orcid{0000-0003-2508-0046}\inst{\ref{aff76}}
\and C.~J.~Conselice\orcid{0000-0003-1949-7638}\inst{\ref{aff16}}
\and L.~Conversi\orcid{0000-0002-6710-8476}\inst{\ref{aff77},\ref{aff52}}
\and Y.~Copin\orcid{0000-0002-5317-7518}\inst{\ref{aff78}}
\and M.~Cropper\orcid{0000-0003-4571-9468}\inst{\ref{aff79}}
\and A.~Da~Silva\orcid{0000-0002-6385-1609}\inst{\ref{aff80},\ref{aff81}}
\and H.~Degaudenzi\orcid{0000-0002-5887-6799}\inst{\ref{aff82}}
\and G.~De~Lucia\orcid{0000-0002-6220-9104}\inst{\ref{aff59}}
\and A.~M.~Di~Giorgio\orcid{0000-0002-4767-2360}\inst{\ref{aff83}}
\and J.~Dinis\orcid{0000-0001-5075-1601}\inst{\ref{aff80},\ref{aff81}}
\and F.~Dubath\orcid{0000-0002-6533-2810}\inst{\ref{aff82}}
\and X.~Dupac\inst{\ref{aff52}}
\and S.~Dusini\orcid{0000-0002-1128-0664}\inst{\ref{aff84}}
\and M.~Farina\orcid{0000-0002-3089-7846}\inst{\ref{aff83}}
\and S.~Farrens\orcid{0000-0002-9594-9387}\inst{\ref{aff85}}
\and F.~Faustini\orcid{0000-0001-6274-5145}\inst{\ref{aff86},\ref{aff73}}
\and S.~Ferriol\inst{\ref{aff78}}
\and M.~Frailis\orcid{0000-0002-7400-2135}\inst{\ref{aff59}}
\and E.~Franceschi\orcid{0000-0002-0585-6591}\inst{\ref{aff13}}
\and S.~Galeotta\orcid{0000-0002-3748-5115}\inst{\ref{aff59}}
\and K.~George\orcid{0000-0002-1734-8455}\inst{\ref{aff29}}
\and W.~Gillard\orcid{0000-0003-4744-9748}\inst{\ref{aff87}}
\and B.~Gillis\orcid{0000-0002-4478-1270}\inst{\ref{aff76}}
\and C.~Giocoli\orcid{0000-0002-9590-7961}\inst{\ref{aff13},\ref{aff28}}
\and P.~G\'omez-Alvarez\orcid{0000-0002-8594-5358}\inst{\ref{aff88},\ref{aff52}}
\and A.~Grazian\orcid{0000-0002-5688-0663}\inst{\ref{aff63}}
\and F.~Grupp\inst{\ref{aff30},\ref{aff29}}
\and S.~V.~H.~Haugan\orcid{0000-0001-9648-7260}\inst{\ref{aff89}}
\and W.~Holmes\inst{\ref{aff90}}
\and I.~Hook\orcid{0000-0002-2960-978X}\inst{\ref{aff91}}
\and F.~Hormuth\inst{\ref{aff92}}
\and A.~Hornstrup\orcid{0000-0002-3363-0936}\inst{\ref{aff93},\ref{aff94}}
\and P.~Hudelot\inst{\ref{aff23}}
\and M.~Jhabvala\inst{\ref{aff95}}
\and B.~Joachimi\orcid{0000-0001-7494-1303}\inst{\ref{aff96}}
\and E.~Keih\"anen\orcid{0000-0003-1804-7715}\inst{\ref{aff97}}
\and S.~Kermiche\orcid{0000-0002-0302-5735}\inst{\ref{aff87}}
\and A.~Kiessling\orcid{0000-0002-2590-1273}\inst{\ref{aff90}}
\and M.~Kilbinger\orcid{0000-0001-9513-7138}\inst{\ref{aff85}}
\and B.~Kubik\orcid{0009-0006-5823-4880}\inst{\ref{aff78}}
\and M.~K\"ummel\orcid{0000-0003-2791-2117}\inst{\ref{aff29}}
\and M.~Kunz\orcid{0000-0002-3052-7394}\inst{\ref{aff98}}
\and H.~Kurki-Suonio\orcid{0000-0002-4618-3063}\inst{\ref{aff99},\ref{aff100}}
\and D.~Le~Mignant\orcid{0000-0002-5339-5515}\inst{\ref{aff22}}
\and S.~Ligori\orcid{0000-0003-4172-4606}\inst{\ref{aff64}}
\and P.~B.~Lilje\orcid{0000-0003-4324-7794}\inst{\ref{aff89}}
\and V.~Lindholm\orcid{0000-0003-2317-5471}\inst{\ref{aff99},\ref{aff100}}
\and I.~Lloro\orcid{0000-0001-5966-1434}\inst{\ref{aff101}}
\and E.~Maiorano\orcid{0000-0003-2593-4355}\inst{\ref{aff13}}
\and O.~Mansutti\orcid{0000-0001-5758-4658}\inst{\ref{aff59}}
\and O.~Marggraf\orcid{0000-0001-7242-3852}\inst{\ref{aff102}}
\and K.~Markovic\orcid{0000-0001-6764-073X}\inst{\ref{aff90}}
\and M.~Martinelli\orcid{0000-0002-6943-7732}\inst{\ref{aff73},\ref{aff103}}
\and N.~Martinet\orcid{0000-0003-2786-7790}\inst{\ref{aff22}}
\and F.~Marulli\orcid{0000-0002-8850-0303}\inst{\ref{aff12},\ref{aff13},\ref{aff28}}
\and R.~Massey\orcid{0000-0002-6085-3780}\inst{\ref{aff37}}
\and E.~Medinaceli\orcid{0000-0002-4040-7783}\inst{\ref{aff13}}
\and S.~Mei\orcid{0000-0002-2849-559X}\inst{\ref{aff104}}
\and M.~Melchior\inst{\ref{aff14}}
\and Y.~Mellier\inst{\ref{aff105},\ref{aff23}}
\and E.~Merlin\orcid{0000-0001-6870-8900}\inst{\ref{aff73}}
\and G.~Meylan\inst{\ref{aff17}}
\and M.~Moresco\orcid{0000-0002-7616-7136}\inst{\ref{aff12},\ref{aff13}}
\and L.~Moscardini\orcid{0000-0002-3473-6716}\inst{\ref{aff12},\ref{aff13},\ref{aff28}}
\and R.~Nakajima\inst{\ref{aff102}}
\and C.~Neissner\orcid{0000-0001-8524-4968}\inst{\ref{aff106},\ref{aff71}}
\and R.~C.~Nichol\orcid{0000-0003-0939-6518}\inst{\ref{aff56}}
\and S.-M.~Niemi\inst{\ref{aff55}}
\and J.~W.~Nightingale\orcid{0000-0002-8987-7401}\inst{\ref{aff107}}
\and C.~Padilla\orcid{0000-0001-7951-0166}\inst{\ref{aff106}}
\and S.~Paltani\orcid{0000-0002-8108-9179}\inst{\ref{aff82}}
\and F.~Pasian\orcid{0000-0002-4869-3227}\inst{\ref{aff59}}
\and K.~Pedersen\inst{\ref{aff108}}
\and W.~J.~Percival\orcid{0000-0002-0644-5727}\inst{\ref{aff109},\ref{aff110},\ref{aff111}}
\and V.~Pettorino\inst{\ref{aff55}}
\and S.~Pires\orcid{0000-0002-0249-2104}\inst{\ref{aff85}}
\and G.~Polenta\orcid{0000-0003-4067-9196}\inst{\ref{aff86}}
\and M.~Poncet\inst{\ref{aff112}}
\and L.~A.~Popa\inst{\ref{aff113}}
\and L.~Pozzetti\orcid{0000-0001-7085-0412}\inst{\ref{aff13}}
\and F.~Raison\orcid{0000-0002-7819-6918}\inst{\ref{aff30}}
\and A.~Renzi\orcid{0000-0001-9856-1970}\inst{\ref{aff114},\ref{aff84}}
\and J.~Rhodes\orcid{0000-0002-4485-8549}\inst{\ref{aff90}}
\and G.~Riccio\inst{\ref{aff3}}
\and E.~Romelli\orcid{0000-0003-3069-9222}\inst{\ref{aff59}}
\and M.~Roncarelli\orcid{0000-0001-9587-7822}\inst{\ref{aff13}}
\and E.~Rossetti\orcid{0000-0003-0238-4047}\inst{\ref{aff62}}
\and R.~Saglia\orcid{0000-0003-0378-7032}\inst{\ref{aff29},\ref{aff30}}
\and Z.~Sakr\orcid{0000-0002-4823-3757}\inst{\ref{aff115},\ref{aff24},\ref{aff116}}
\and A.~G.~S\'anchez\orcid{0000-0003-1198-831X}\inst{\ref{aff30}}
\and D.~Sapone\orcid{0000-0001-7089-4503}\inst{\ref{aff117}}
\and B.~Sartoris\orcid{0000-0003-1337-5269}\inst{\ref{aff29},\ref{aff59}}
\and P.~Schneider\orcid{0000-0001-8561-2679}\inst{\ref{aff102}}
\and T.~Schrabback\orcid{0000-0002-6987-7834}\inst{\ref{aff118}}
\and A.~Secroun\orcid{0000-0003-0505-3710}\inst{\ref{aff87}}
\and G.~Seidel\orcid{0000-0003-2907-353X}\inst{\ref{aff35}}
\and S.~Serrano\orcid{0000-0002-0211-2861}\inst{\ref{aff119},\ref{aff120},\ref{aff121}}
\and C.~Sirignano\orcid{0000-0002-0995-7146}\inst{\ref{aff114},\ref{aff84}}
\and G.~Sirri\orcid{0000-0003-2626-2853}\inst{\ref{aff28}}
\and J.~Skottfelt\orcid{0000-0003-1310-8283}\inst{\ref{aff122}}
\and L.~Stanco\orcid{0000-0002-9706-5104}\inst{\ref{aff84}}
\and J.~Steinwagner\orcid{0000-0001-7443-1047}\inst{\ref{aff30}}
\and P.~Tallada-Cresp\'{i}\orcid{0000-0002-1336-8328}\inst{\ref{aff70},\ref{aff71}}
\and I.~Tereno\inst{\ref{aff80},\ref{aff123}}
\and R.~Toledo-Moreo\orcid{0000-0002-2997-4859}\inst{\ref{aff124}}
\and F.~Torradeflot\orcid{0000-0003-1160-1517}\inst{\ref{aff71},\ref{aff70}}
\and I.~Tutusaus\orcid{0000-0002-3199-0399}\inst{\ref{aff24}}
\and E.~A.~Valentijn\inst{\ref{aff2}}
\and L.~Valenziano\orcid{0000-0002-1170-0104}\inst{\ref{aff13},\ref{aff125}}
\and T.~Vassallo\orcid{0000-0001-6512-6358}\inst{\ref{aff29},\ref{aff59}}
\and G.~Verdoes~Kleijn\orcid{0000-0001-5803-2580}\inst{\ref{aff2}}
\and A.~Veropalumbo\orcid{0000-0003-2387-1194}\inst{\ref{aff57},\ref{aff66},\ref{aff126}}
\and Y.~Wang\orcid{0000-0002-4749-2984}\inst{\ref{aff127}}
\and J.~Weller\orcid{0000-0002-8282-2010}\inst{\ref{aff29},\ref{aff30}}
\and G.~Zamorani\orcid{0000-0002-2318-301X}\inst{\ref{aff13}}
\and E.~Zucca\orcid{0000-0002-5845-8132}\inst{\ref{aff13}}
\and C.~Burigana\orcid{0000-0002-3005-5796}\inst{\ref{aff39},\ref{aff125}}
\and M.~Calabrese\orcid{0000-0002-2637-2422}\inst{\ref{aff128},\ref{aff42}}
\and A.~Mora\orcid{0000-0002-1922-8529}\inst{\ref{aff129}}
\and M.~P\"ontinen\orcid{0000-0001-5442-2530}\inst{\ref{aff99}}
\and V.~Scottez\inst{\ref{aff105},\ref{aff130}}
\and M.~Viel\orcid{0000-0002-2642-5707}\inst{\ref{aff58},\ref{aff59},\ref{aff61},\ref{aff60},\ref{aff131}}
\and B.~Margalef-Bentabol\orcid{0000-0001-8702-7019}\inst{\ref{aff132}}}
										   
\institute{School of Physical Sciences, The Open University, Milton Keynes, MK7 6AA, UK\label{aff1}
\and
Kapteyn Astronomical Institute, University of Groningen, PO Box 800, 9700 AV Groningen, The Netherlands\label{aff2}
\and
INAF-Osservatorio Astronomico di Capodimonte, Via Moiariello 16, 80131 Napoli, Italy\label{aff3}
\and
Department of Physics "E. Pancini", University Federico II, Via Cinthia 6, 80126, Napoli, Italy\label{aff4}
\and
INFN section of Naples, Via Cinthia 6, 80126, Napoli, Italy\label{aff5}
\and
University of Trento, Via Sommarive 14, I-38123 Trento, Italy\label{aff6}
\and
Dipartimento di Fisica e Astronomia, Universit\`{a} di Firenze, via G. Sansone 1, 50019 Sesto Fiorentino, Firenze, Italy\label{aff7}
\and
INAF-Osservatorio Astrofisico di Arcetri, Largo E. Fermi 5, 50125, Firenze, Italy\label{aff8}
\and
Technical University of Munich, TUM School of Natural Sciences, Physics Department, James-Franck-Str.~1, 85748 Garching, Germany\label{aff9}
\and
Max-Planck-Institut f\"ur Astrophysik, Karl-Schwarzschild-Str.~1, 85748 Garching, Germany\label{aff10}
\and
Departamento F\'isica Aplicada, Universidad Polit\'ecnica de Cartagena, Campus Muralla del Mar, 30202 Cartagena, Murcia, Spain\label{aff11}
\and
Dipartimento di Fisica e Astronomia "Augusto Righi" - Alma Mater Studiorum Universit\`a di Bologna, via Piero Gobetti 93/2, 40129 Bologna, Italy\label{aff12}
\and
INAF-Osservatorio di Astrofisica e Scienza dello Spazio di Bologna, Via Piero Gobetti 93/3, 40129 Bologna, Italy\label{aff13}
\and
University of Applied Sciences and Arts of Northwestern Switzerland, School of Engineering, 5210 Windisch, Switzerland\label{aff14}
\and
David A. Dunlap Department of Astronomy \& Astrophysics, University of Toronto, 50 St George Street, Toronto, Ontario M5S 3H4, Canada\label{aff15}
\and
Jodrell Bank Centre for Astrophysics, Department of Physics and Astronomy, University of Manchester, Oxford Road, Manchester M13 9PL, UK\label{aff16}
\and
Institute of Physics, Laboratory of Astrophysics, Ecole Polytechnique F\'ed\'erale de Lausanne (EPFL), Observatoire de Sauverny, 1290 Versoix, Switzerland\label{aff17}
\and
SCITAS, Ecole Polytechnique F\'ed\'erale de Lausanne (EPFL), 1015 Lausanne, Switzerland\label{aff18}
\and
Institute of Cosmology and Gravitation, University of Portsmouth, Portsmouth PO1 3FX, UK\label{aff19}
\and
Institut de Ci\`{e}ncies del Cosmos (ICCUB), Universitat de Barcelona (IEEC-UB), Mart\'{i} i Franqu\`{e}s 1, 08028 Barcelona, Spain\label{aff20}
\and
Instituci\'o Catalana de Recerca i Estudis Avan\c{c}ats (ICREA), Passeig de Llu\'{\i}s Companys 23, 08010 Barcelona, Spain\label{aff21}
\and
Aix-Marseille Universit\'e, CNRS, CNES, LAM, Marseille, France\label{aff22}
\and
Institut d'Astrophysique de Paris, UMR 7095, CNRS, and Sorbonne Universit\'e, 98 bis boulevard Arago, 75014 Paris, France\label{aff23}
\and
Institut de Recherche en Astrophysique et Plan\'etologie (IRAP), Universit\'e de Toulouse, CNRS, UPS, CNES, 14 Av. Edouard Belin, 31400 Toulouse, France\label{aff24}
\and
UCB Lyon 1, CNRS/IN2P3, IUF, IP2I Lyon, 4 rue Enrico Fermi, 69622 Villeurbanne, France\label{aff25}
\and
Department of Astronomy, University of Cape Town, Rondebosch, Cape Town, 7700, South Africa\label{aff26}
\and
STAR Institute, Quartier Agora - All\'ee du six Ao\^ut, 19c B-4000 Li\`ege, Belgium\label{aff27}
\and
INFN-Sezione di Bologna, Viale Berti Pichat 6/2, 40127 Bologna, Italy\label{aff28}
\and
Universit\"ats-Sternwarte M\"unchen, Fakult\"at f\"ur Physik, Ludwig-Maximilians-Universit\"at M\"unchen, Scheinerstrasse 1, 81679 M\"unchen, Germany\label{aff29}
\and
Max Planck Institute for Extraterrestrial Physics, Giessenbachstr. 1, 85748 Garching, Germany\label{aff30}
\and
INFN, Sezione di Lecce, Via per Arnesano, CP-193, 73100, Lecce, Italy\label{aff31}
\and
Department of Mathematics and Physics E. De Giorgi, University of Salento, Via per Arnesano, CP-I93, 73100, Lecce, Italy\label{aff32}
\and
INAF-Sezione di Lecce, c/o Dipartimento Matematica e Fisica, Via per Arnesano, 73100, Lecce, Italy\label{aff33}
\and
Department of Physics, Oxford University, Keble Road, Oxford OX1 3RH, UK\label{aff34}
\and
Max-Planck-Institut f\"ur Astronomie, K\"onigstuhl 17, 69117 Heidelberg, Germany\label{aff35}
\and
Department of Physics, Centre for Extragalactic Astronomy, Durham University, South Road, Durham, DH1 3LE, UK\label{aff36}
\and
Department of Physics, Institute for Computational Cosmology, Durham University, South Road, Durham, DH1 3LE, UK\label{aff37}
\and
Inter-University Institute for Data Intensive Astronomy, Department of Astronomy, University of Cape Town, 7701 Rondebosch, Cape Town, South Africa\label{aff38}
\and
INAF, Istituto di Radioastronomia, Via Piero Gobetti 101, 40129 Bologna, Italy\label{aff39}
\and
Minnesota Institute for Astrophysics, University of Minnesota, 116 Church St SE, Minneapolis, MN 55455, USA\label{aff40}
\and
Dipartimento di Fisica "Aldo Pontremoli", Universit\`a degli Studi di Milano, Via Celoria 16, 20133 Milano, Italy\label{aff41}
\and
INAF-IASF Milano, Via Alfonso Corti 12, 20133 Milano, Italy\label{aff42}
\and
Inter-University Institute for Data Intensive Astronomy, Department of Physics and Astronomy, University of the Western Cape, 7535 Bellville, Cape Town, South Africa\label{aff43}
\and
Department of Physics and Astronomy, Lehman College of the CUNY, Bronx, NY 10468, USA\label{aff44}
\and
American Museum of Natural History, Department of Astrophysics, New York, NY 10024, USA\label{aff45}
\and
Universit\'e de Strasbourg, CNRS, Observatoire astronomique de Strasbourg, UMR 7550, 67000 Strasbourg, France\label{aff46}
\and
Department of Physics, Universit\'{e} de Montr\'{e}al, 2900 Edouard Montpetit Blvd, Montr\'{e}al, Qu\'{e}bec H3T 1J4, Canada\label{aff47}
\and
Centro de Astrof\'{\i}sica da Universidade do Porto, Rua das Estrelas, 4150-762 Porto, Portugal\label{aff48}
\and
Instituto de Astrof\'isica e Ci\^encias do Espa\c{c}o, Universidade do Porto, CAUP, Rua das Estrelas, PT4150-762 Porto, Portugal\label{aff49}
\and
Dipartimento di Fisica e Scienze della Terra, Universit\`a degli Studi di Ferrara, Via Giuseppe Saragat 1, 44122 Ferrara, Italy\label{aff50}
\and
Laboratoire univers et particules de Montpellier, Universit\'e de Montpellier, CNRS, 34090 Montpellier, France\label{aff51}
\and
ESAC/ESA, Camino Bajo del Castillo, s/n., Urb. Villafranca del Castillo, 28692 Villanueva de la Ca\~nada, Madrid, Spain\label{aff52}
\and
School of Computing and Communications, The Open University, Milton Keynes, MK7 6AA, UK\label{aff53}
\and
Laboratoire d'Astrophysique de Bordeaux, CNRS and Universit\'e de Bordeaux, All\'ee Geoffroy St. Hilaire, 33165 Pessac, France\label{aff54}
\and
European Space Agency/ESTEC, Keplerlaan 1, 2201 AZ Noordwijk, The Netherlands\label{aff55}
\and
School of Mathematics and Physics, University of Surrey, Guildford, Surrey, GU2 7XH, UK\label{aff56}
\and
INAF-Osservatorio Astronomico di Brera, Via Brera 28, 20122 Milano, Italy\label{aff57}
\and
IFPU, Institute for Fundamental Physics of the Universe, via Beirut 2, 34151 Trieste, Italy\label{aff58}
\and
INAF-Osservatorio Astronomico di Trieste, Via G. B. Tiepolo 11, 34143 Trieste, Italy\label{aff59}
\and
INFN, Sezione di Trieste, Via Valerio 2, 34127 Trieste TS, Italy\label{aff60}
\and
SISSA, International School for Advanced Studies, Via Bonomea 265, 34136 Trieste TS, Italy\label{aff61}
\and
Dipartimento di Fisica e Astronomia, Universit\`a di Bologna, Via Gobetti 93/2, 40129 Bologna, Italy\label{aff62}
\and
INAF-Osservatorio Astronomico di Padova, Via dell'Osservatorio 5, 35122 Padova, Italy\label{aff63}
\and
INAF-Osservatorio Astrofisico di Torino, Via Osservatorio 20, 10025 Pino Torinese (TO), Italy\label{aff64}
\and
Dipartimento di Fisica, Universit\`a di Genova, Via Dodecaneso 33, 16146, Genova, Italy\label{aff65}
\and
INFN-Sezione di Genova, Via Dodecaneso 33, 16146, Genova, Italy\label{aff66}
\and
Faculdade de Ci\^encias da Universidade do Porto, Rua do Campo de Alegre, 4150-007 Porto, Portugal\label{aff67}
\and
Dipartimento di Fisica, Universit\`a degli Studi di Torino, Via P. Giuria 1, 10125 Torino, Italy\label{aff68}
\and
INFN-Sezione di Torino, Via P. Giuria 1, 10125 Torino, Italy\label{aff69}
\and
Centro de Investigaciones Energ\'eticas, Medioambientales y Tecnol\'ogicas (CIEMAT), Avenida Complutense 40, 28040 Madrid, Spain\label{aff70}
\and
Port d'Informaci\'{o} Cient\'{i}fica, Campus UAB, C. Albareda s/n, 08193 Bellaterra (Barcelona), Spain\label{aff71}
\and
Institute for Theoretical Particle Physics and Cosmology (TTK), RWTH Aachen University, 52056 Aachen, Germany\label{aff72}
\and
INAF-Osservatorio Astronomico di Roma, Via Frascati 33, 00078 Monteporzio Catone, Italy\label{aff73}
\and
Dipartimento di Fisica e Astronomia "Augusto Righi" - Alma Mater Studiorum Universit\`a di Bologna, Viale Berti Pichat 6/2, 40127 Bologna, Italy\label{aff74}
\and
Instituto de Astrof\'isica de Canarias, Calle V\'ia L\'actea s/n, 38204, San Crist\'obal de La Laguna, Tenerife, Spain\label{aff75}
\and
Institute for Astronomy, University of Edinburgh, Royal Observatory, Blackford Hill, Edinburgh EH9 3HJ, UK\label{aff76}
\and
European Space Agency/ESRIN, Largo Galileo Galilei 1, 00044 Frascati, Roma, Italy\label{aff77}
\and
Universit\'e Claude Bernard Lyon 1, CNRS/IN2P3, IP2I Lyon, UMR 5822, Villeurbanne, F-69100, France\label{aff78}
\and
Mullard Space Science Laboratory, University College London, Holmbury St Mary, Dorking, Surrey RH5 6NT, UK\label{aff79}
\and
Departamento de F\'isica, Faculdade de Ci\^encias, Universidade de Lisboa, Edif\'icio C8, Campo Grande, PT1749-016 Lisboa, Portugal\label{aff80}
\and
Instituto de Astrof\'isica e Ci\^encias do Espa\c{c}o, Faculdade de Ci\^encias, Universidade de Lisboa, Campo Grande, 1749-016 Lisboa, Portugal\label{aff81}
\and
Department of Astronomy, University of Geneva, ch. d'Ecogia 16, 1290 Versoix, Switzerland\label{aff82}
\and
INAF-Istituto di Astrofisica e Planetologia Spaziali, via del Fosso del Cavaliere, 100, 00100 Roma, Italy\label{aff83}
\and
INFN-Padova, Via Marzolo 8, 35131 Padova, Italy\label{aff84}
\and
Universit\'e Paris-Saclay, Universit\'e Paris Cit\'e, CEA, CNRS, AIM, 91191, Gif-sur-Yvette, France\label{aff85}
\and
Space Science Data Center, Italian Space Agency, via del Politecnico snc, 00133 Roma, Italy\label{aff86}
\and
Aix-Marseille Universit\'e, CNRS/IN2P3, CPPM, Marseille, France\label{aff87}
\and
FRACTAL S.L.N.E., calle Tulip\'an 2, Portal 13 1A, 28231, Las Rozas de Madrid, Spain\label{aff88}
\and
Institute of Theoretical Astrophysics, University of Oslo, P.O. Box 1029 Blindern, 0315 Oslo, Norway\label{aff89}
\and
Jet Propulsion Laboratory, California Institute of Technology, 4800 Oak Grove Drive, Pasadena, CA, 91109, USA\label{aff90}
\and
Department of Physics, Lancaster University, Lancaster, LA1 4YB, UK\label{aff91}
\and
Felix Hormuth Engineering, Goethestr. 17, 69181 Leimen, Germany\label{aff92}
\and
Technical University of Denmark, Elektrovej 327, 2800 Kgs. Lyngby, Denmark\label{aff93}
\and
Cosmic Dawn Center (DAWN), Denmark\label{aff94}
\and
NASA Goddard Space Flight Center, Greenbelt, MD 20771, USA\label{aff95}
\and
Department of Physics and Astronomy, University College London, Gower Street, London WC1E 6BT, UK\label{aff96}
\and
Department of Physics and Helsinki Institute of Physics, Gustaf H\"allstr\"omin katu 2, 00014 University of Helsinki, Finland\label{aff97}
\and
Universit\'e de Gen\`eve, D\'epartement de Physique Th\'eorique and Centre for Astroparticle Physics, 24 quai Ernest-Ansermet, CH-1211 Gen\`eve 4, Switzerland\label{aff98}
\and
Department of Physics, P.O. Box 64, 00014 University of Helsinki, Finland\label{aff99}
\and
Helsinki Institute of Physics, Gustaf H{\"a}llstr{\"o}min katu 2, University of Helsinki, Helsinki, Finland\label{aff100}
\and
NOVA optical infrared instrumentation group at ASTRON, Oude Hoogeveensedijk 4, 7991PD, Dwingeloo, The Netherlands\label{aff101}
\and
Universit\"at Bonn, Argelander-Institut f\"ur Astronomie, Auf dem H\"ugel 71, 53121 Bonn, Germany\label{aff102}
\and
INFN-Sezione di Roma, Piazzale Aldo Moro, 2 - c/o Dipartimento di Fisica, Edificio G. Marconi, 00185 Roma, Italy\label{aff103}
\and
Universit\'e Paris Cit\'e, CNRS, Astroparticule et Cosmologie, 75013 Paris, France\label{aff104}
\and
Institut d'Astrophysique de Paris, 98bis Boulevard Arago, 75014, Paris, France\label{aff105}
\and
Institut de F\'{i}sica d'Altes Energies (IFAE), The Barcelona Institute of Science and Technology, Campus UAB, 08193 Bellaterra (Barcelona), Spain\label{aff106}
\and
School of Mathematics, Statistics and Physics, Newcastle University, Herschel Building, Newcastle-upon-Tyne, NE1 7RU, UK\label{aff107}
\and
DARK, Niels Bohr Institute, University of Copenhagen, Jagtvej 155, 2200 Copenhagen, Denmark\label{aff108}
\and
Waterloo Centre for Astrophysics, University of Waterloo, Waterloo, Ontario N2L 3G1, Canada\label{aff109}
\and
Department of Physics and Astronomy, University of Waterloo, Waterloo, Ontario N2L 3G1, Canada\label{aff110}
\and
Perimeter Institute for Theoretical Physics, Waterloo, Ontario N2L 2Y5, Canada\label{aff111}
\and
Centre National d'Etudes Spatiales -- Centre spatial de Toulouse, 18 avenue Edouard Belin, 31401 Toulouse Cedex 9, France\label{aff112}
\and
Institute of Space Science, Str. Atomistilor, nr. 409 M\u{a}gurele, Ilfov, 077125, Romania\label{aff113}
\and
Dipartimento di Fisica e Astronomia "G. Galilei", Universit\`a di Padova, Via Marzolo 8, 35131 Padova, Italy\label{aff114}
\and
Institut f\"ur Theoretische Physik, University of Heidelberg, Philosophenweg 16, 69120 Heidelberg, Germany\label{aff115}
\and
Universit\'e St Joseph; Faculty of Sciences, Beirut, Lebanon\label{aff116}
\and
Departamento de F\'isica, FCFM, Universidad de Chile, Blanco Encalada 2008, Santiago, Chile\label{aff117}
\and
Universit\"at Innsbruck, Institut f\"ur Astro- und Teilchenphysik, Technikerstr. 25/8, 6020 Innsbruck, Austria\label{aff118}
\and
Institut d'Estudis Espacials de Catalunya (IEEC),  Edifici RDIT, Campus UPC, 08860 Castelldefels, Barcelona, Spain\label{aff119}
\and
Satlantis, University Science Park, Sede Bld 48940, Leioa-Bilbao, Spain\label{aff120}
\and
Institute of Space Sciences (ICE, CSIC), Campus UAB, Carrer de Can Magrans, s/n, 08193 Barcelona, Spain\label{aff121}
\and
Centre for Electronic Imaging, Open University, Walton Hall, Milton Keynes, MK7~6AA, UK\label{aff122}
\and
Instituto de Astrof\'isica e Ci\^encias do Espa\c{c}o, Faculdade de Ci\^encias, Universidade de Lisboa, Tapada da Ajuda, 1349-018 Lisboa, Portugal\label{aff123}
\and
Universidad Polit\'ecnica de Cartagena, Departamento de Electr\'onica y Tecnolog\'ia de Computadoras,  Plaza del Hospital 1, 30202 Cartagena, Spain\label{aff124}
\and
INFN-Bologna, Via Irnerio 46, 40126 Bologna, Italy\label{aff125}
\and
Dipartimento di Fisica, Universit\`a degli studi di Genova, and INFN-Sezione di Genova, via Dodecaneso 33, 16146, Genova, Italy\label{aff126}
\and
Infrared Processing and Analysis Center, California Institute of Technology, Pasadena, CA 91125, USA\label{aff127}
\and
Astronomical Observatory of the Autonomous Region of the Aosta Valley (OAVdA), Loc. Lignan 39, I-11020, Nus (Aosta Valley), Italy\label{aff128}
\and
Aurora Technology for European Space Agency (ESA), Camino bajo del Castillo, s/n, Urbanizacion Villafranca del Castillo, Villanueva de la Ca\~nada, 28692 Madrid, Spain\label{aff129}
\and
ICL, Junia, Universit\'e Catholique de Lille, LITL, 59000 Lille, France\label{aff130}
\and
ICSC - Centro Nazionale di Ricerca in High Performance Computing, Big Data e Quantum Computing, Via Magnanelli 2, Bologna, Italy\label{aff131}
\and
SRON Netherlands Institute for Space Research, Landleven 12, 9747 AD, Groningen, The Netherlands\label{aff132}}    	 

										   

 \date{\today}

%
%
 \abstract{The Euclid Wide Survey (EWS) is predicted to find approximately \num{170000} galaxy-galaxy strong lenses from its lifetime observation of $\num{14000}\textrm{\,deg}^2$ of the sky. Detecting this many lenses by visual inspection with professional astronomers and citizen scientists alone is infeasible. As a result, machine learning algorithms, particularly convolutional neural networks (CNNs), have been used as an automated method of detecting strong lenses, and have proven fruitful in finding galaxy-galaxy strong lens candidates, such that the usage of CNNs in lens identification has increased. We identify the major challenge to be the automatic detection of galaxy-galaxy strong lenses while simultaneously maintaining a low false positive rate, thus producing a pure and complete sample of strong lens candidates from \Euclid with a limited need for visual inspection. 
 One aim of this research is to have a quantified starting point on the achieved purity and completeness with our current version of CNN-based detection pipelines for the VIS images of EWS. This work is vital in preparing our CNN-based detection pipelines to be able to produce a pure sample of the $>\num{100000}$ strong gravitational lensing systems widely predicted for \Euclid.
 We select all sources with VIS  $\IE<23$ mag from the \Euclid Early Release Observation imaging of the Perseus field. We apply a range of CNN architectures to detect strong lenses in these cutouts. All our networks perform extremely well on simulated data sets and their respective validation sets. However, when applied to real \Euclid imaging, the highest lens purity is just $\sim11$\%. Among all our networks, the false positives are typically identifiable by human volunteers as, for example, spiral galaxies, multiple sources, and artifacts, implying that improvements are still possible, perhaps via a second, more interpretable lens selection filtering stage. There is currently no alternative to human classification of CNN-selected lens candidates. Given the expected $\sim10^5$ lensing systems in \Euclid, this implies $10^6$ objects for human classification, which while very large is not in principle intractable and not without precedent.
 }
%
%
%
%
   \titlerunning{\Euclid\/: Searches for strong gravitational lenses}
   \authorrunning{R. Pearce-Casey et al.}
   
   \maketitle
%
%
%
%
   
\section{\label{sc:Intro}Introduction}
\Euclid \citep{laureijs2011euclid,EuclidSkyOverview} is a European Space Agency (ESA) medium-class mission with the main objectives to study dark matter and dark energy, the dominant yet elusive 
components of the Universe, and address the tensions that have recently emerged within the standard cosmological model (\citealp{Ilic2022EuclidAndCMB}). \Euclid maps the evolution of mass distribution throughout cosmic time using weak lensing, baryonic acoustic oscillations, and the large-scale structure.

Strong gravitational lensing is a phenomenon where the light from a distant source, or multiple sources, is deflected by a massive foreground object, 
which results in the creation of distinct multiple images, arcs, or rings, depending on the nature of the source and the alignment. Strong lensing has several applications such as providing constraints on dark energy (\citealp{Dark_Energy_4}; \citealp{Dark_Energy_3}; \citealp{Dark_Energy_2}; \citealp{Dark_Energy_5}; \citealp{Dark_Energy_1}), the mass distribution of galaxies (\citealp{1991ApJ...370....1M}; \citealp{moller2002probing}; \citealp{limousin2005constraining}), dark matter (\citealp{vegetti2023stronggravitationallensingprobe}; \citealp{Powell_2023}), and the slope of inner mass density profile (\citealp{IMD_2002}; \citealp{ISD_2009}; \citealp{IMD_2012}; \citealp{ICD_2015b}; \citealp{ISD_2015}), which are crucial 
legacy science 
objectives of the \Euclid mission.

Galaxy-scale strong lenses are rare; often only 
one in \mbox{$\approx10^3$-$10^4$} 
galaxies in blank-field surveys are
identified as being part of strong gravitational lensing systems (\citealp{100000_lenses_3}). To date, 
around
a thousand gravitational lenses have been found across many heterogeneous data sets (e.g. \citealp{Lehar_2000,Nord_2016,Shu_2017,Petrillo_2}). {ESA\textquotesingle}s \Euclid space telescope is expected to increase the number of potential strong gravitational lens candidates by orders of magnitude (\citealp{100000_lenses_3}). For example, it is estimated that there will be approximately \num{170000} observable galaxy-galaxy lenses in the \Euclid data set among over a billion potential objects (\citealp{SGL_CNN_app_2019c}). This will bring about a new era for strong lensing where large relatively well-defined samples of lenses will be available. \Euclid has already started to find strong gravitational lenses (e.g. \citealp{barroso2024euclidearlyreleaseobservations}; \textcolor{blue}{O'Riordan et al. 2024} in prep). While it is easy, in principle, to create a highly complete catalogue of strong lensing systems from \Euclid, it is not at all obvious how to make this sample reliable. The most commonly used method for finding lenses in imaging surveys has been by visual inspection of candidates selected on the basis of luminosity and/or colour. However, the scale of \Euclid's data set makes classification with human volunteers alone near to impossible. We therefore require an automated approach, a task that is well suited to using machine learning.

Owing to their success in identifying strong lenses in the strong gravitational lens finding challenge (\citealp{SGL_CNN_app_2019c}), convolutional 
neural networks (CNNs, \citealp{6795724}) have become widely preferred over other machine learning techniques. CNNs excel at feature learning from imaging data, and the spatial information extracted makes it useful for understanding the image content. To learn features, CNNs generally use several convolutional and pooling layers followed by fully-connected layers in order to make a prediction. CNNs are trained by associating the given labels with the extracted features through gradient-based back-propagation (\citealp{hecht1992theory}; \citealp{buscema1998back}).
   
CNNs are a widely-used technique
for
identifying strong gravitational lenses 
across many recent surveys
(\citealp{Petrillo_1, Petrillo_2, Petrillo_3}; \citealp{lanusse2018cmu}; \citealp{SGL_CNN_app_pour}; \citealp{schaefer2018deep}; \citealp{SGL_CNN_app_2019b}; \citealp{SGL_CNN_app_2019c};  \citealp{SGL_CNN_app_2020, SGL_CNN_app_2021}; \citealp{canameras2020holismokes}; \citealp{SGL_CNN_app_christ}; \citealp{2022MNRAS.510..500G}; \citealp{Rezaei}; \citealp{Andika_2023}, and \citealp{nagam}). The most popular architectures that have been used in finding strong lenses are \texttt{EfficientNet} (\citealp{pmlr-v97-tan19a}), \texttt{DenseNets} (\citealp{huang2017densely}; \citealp[e.g.][]{nagam,nagam2}), and \texttt{ResNet} (\citealp{he_2016}; \citealp[e.g.][]{Thuruthipilly_2022,storfer2023newstronggravitationallenses}).
   
Here, we perform a systematic comparison of 
networks and training data sets, 
all of which are tested on a common test set. Such a study is 
a powerful diagnostic 
for identifying the strengths and weaknesses of different network architectures and the construction strategies of different training and validation data sets, 
and will inform 
the development of a robust, automated approach to produce a highly efficient lens search. Similar work by \cite{2024MNRAS.tmp.1568M} also shows a comparison and benchmarking of different networks trained on their individual training data sets. The network architectures they implement are described in Sect.\,2 of \cite{2024MNRAS.tmp.1568M}, and involve similar architectures to the networks we use in this work, which we describe in Sect.\,\ref{sec: Methods}. They show that each network performs well on their own constructed test data sets compared to those from others, yet all networks perform comparably on a common test data set of (real) lenses and non-lenses. Similar to their work, we compare our networks, trained and validated on their own respective data sets, and tested on a common test set, which is, in this case, composed of real \Euclid imaging, in which the point spread function (PSF) is sufficiently different to the analysis in \cite{2024MNRAS.tmp.1568M} that the present study is warranted.

The paper is structured as follows. In Sect.\,\ref{sec: Datasets}, we briefly introduce \Euclid's 
Early Release Observation
data and describe the common test data set. In Sect.\,\ref{sec: Methods}, we describe the various networks and the methodologies used in generating the training and validation sets. In Sect.\,\ref{sec:metrics}, we list the performance metrics used in evaluating the various networks, and we present the results in Sect.\,\ref{sec: Results}. We give our conclusions in Sect.\,\ref{sec: Conclusions}.

\section{Early Release Observations}\label{sec: Datasets}
Below, we give a brief overview of the \Euclid space telescope and its Early Release 
Observation (hereafter ERO) data (\citealp{EROcite}). The ERO programme was an initiative of ESA and the Euclid Science Team, including 24 hours of observations that were taken before the start of \Euclid's nominal survey. These ERO data were not part of the nominal survey, and address legacy science rather than \Euclid core science. The test set that we use for the various networks was constructed from one of the six selected ERO proposals where a visual inspection of objects was performed by a team of Euclid Consortium members to produce a truth set of lens candidates for this study. We discuss the methods used to construct this test set here.

\subsection{\Euclid space telescope}
The \Euclid space telescope consists of two major components, namely the service module and the payload module. The service module takes care of the power generation, navigation, and communication modules. The payload module of \Euclid is equipped with two cameras, namely the Visual Imager (VIS, \citealp{EuclidSkyVIS}) and the Near Infrared Spectrometer and Photometer (NISP, \citealp{EuclidSkyNISP}). The VIS camera is equipped with a single broadband filter, operating between wavelengths of 530\,nm and 920\,nm. Additionally, the \Euclid NISP instrument provides multiband photometry in three near-infrared bands, \YE, \JE, and \HE, operating in the 950-2020\,nm wavelength range (\citealp{2022A&A...662A..92E}).

\subsection{Data sets}

The ERO programme committee selected six project proposals covering a total of $17$ ERO fields. \Euclid first released its ERO (\citealp{EROData}) data internally for its Euclid Consortium members. One among the six 
projects 
was to observe a cluster of galaxies and as a result the Perseus cluster of galaxies was 
selected 
to be observed. For this \Euclid ERO programme, $0.7\textrm{\,deg}^2$ of deep observations of the central region of the Perseus cluster were carried out using the broad filter in the optical band \IE with the VIS instrument, and three broad near-infrared filters \YE, \JE, and \HE with the NISP instrument. \cite{cuillandre2024euclidearlyreleaseobservations} shows that there are \num{\sim{1000}} foreground galaxies belonging to the Perseus cluster and \num{100000} distant galaxies occupying the background.

The Euclid Wide Survey (EWS) is designed to image each point of the sky using a single Reference Observing Sequence (ROS; \citealp{Scaramella-EP1}) in which the observations are carried out with four dithered exposures of 87.2 seconds in each of the \YE, \JE, and \HE filters, and four dithered exposures of 566 seconds in \IE for the ERO data of the Perseus cluster.  
The \Euclid performance verification phase was carried out in September 2023 (\citealp{EROData}), separate but in conjunction with \Euclid ERO data. The Perseus cluster ERO data consisted of four ROS with a total integration time of 7456 seconds in the \IE filter, and 1392.2 seconds in the \YE, \JE, and \HE filters thereby achieving a depth 0.75 magnitudes deeper than the EWS.
The Perseus cluster imaging represents an appropriate data set to perform a 
search for gravitational lens systems, and to prepare for future systematic investigations with the EWS data.

\subsection{Construction of the common test set}
The ERO data of the Perseus cluster were reserved for testing and comparing the different networks discussed in this work. A team of Euclid Consortium members performed a systematic visual inspection of the $0.7\textrm{\,deg}^2$ \Euclid ERO data of the Perseus cluster using both the high-resolution VIS \IE-band, and the lower resolution NISP bands (\citealp{barroso2024euclidearlyreleaseobservations}). Every extended source brighter than magnitude $23$ in \IE was inspected, which amounted to $\num{12086}$ stamps of $10\arcsecond\times 10\arcsecond$ with $41$ expert human classifiers.

The detailed classification scheme for classifying the sources (see \citealp{barroso2024euclidearlyreleaseobservations} for details) into one of the following non-overlapping categories, in which $A$, $B$, and $C$ are positive grades, whereas $X$, and $I$ are negative grades, is given as follows:
\begin{itemize}
\item class \textbf{A} represents 
definite
lenses showing clear lensing features; 
\item class \textbf{B} indicates that lensing features are present, but cannot be confirmed without additional information; 
\item class \textbf{C} denotes that lensing features are present, but can also be explained by other physical phenomena; 
\item class \textbf{X} denotes a  
definite non-lens; 
\item class \textbf{I} are interesting objects, but are definitely not
lensing systems.
\end{itemize}

\begin{figure}[t]%
    \centering
    {{\includegraphics[width=9cm]{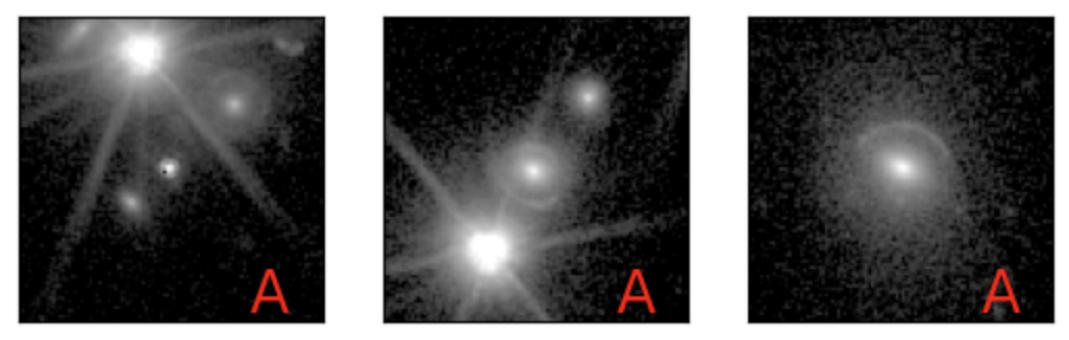} }}
    \caption{The three grade $A$ candidate lenses found by a visual inspection of $\num{12086}$ stamps from the Perseus cluster. The grade $A$ candidates show clear lensing features. The \IE-band images are $10^{\prime\prime}\times10^{\prime\prime}$. Detailed lens models for these candidates are shown in Appendix C of \cite{barroso2024euclidearlyreleaseobservations}.}
    \label{fig:grade a}
\end{figure}

The visual inspection resulted in the discovery of three grade $A$ candidates, $13$ grade $B$ candidates, and $52$ grade $C$ candidates. The $16$ combined grade $A$ and grade $B$ candidates were modelled to assess their validity by checking that they were consistent
with a single source lensed by a plausible mass distribution, for which five candidates passed this test. The three grade $A$ candidates are shown in Fig.\,\ref{fig:grade a}.

For this work, however, the decision was made to label the $68$ combined grade $A$, grade $B$, and grade $C$ candidates as potential lenses. Grade $B$ and $C$ candidates are shown in Fig.\,\ref{fig:ABC}. This corresponds to asking each network to select visually plausible lens candidates. Thus, the common test set to evaluate the various networks is composed of the $\num{12086}$ stamps, inclusive of the $68$ lens candidates being the positives, and the remaining stamps the negatives. 

\section{Overview of networks}\label{sec: Methods}
Here, we give an overview of the various networks with their respective training data sets. The networks described in this section represent our current CNN-based detection methods for finding strong lenses in \Euclid. In total, 20 networks were used, of which six networks were trained with data from the same simulation sets, created using halo catalogues provided by \cite{EuclidSkyFlagship}, though  
these training sets 
differ in size for each network. The remaining networks were trained on their respective training data sets, created specifically for this study using either Galaxy Zoo (\citealp{Masters_2019}, hereafter GZ) and/or the Kilo Degree Survey (KiDS, \citealp{Kids}) data, 
which are of different sizes. The training data sets for each network are described in the following subsections. We present the results of evaluating the 20 networks on the common test set in Sect.\,\ref{sec: Results}, illustrating which networks are successful lens finders, and 
which are therefore good candidates 
for our automated detection pipeline.

\subsection{Zoobot}\label{ssec:Zoobot}
\texttt{Zoobot} \citep{2023JOSS....8.5312W} is a Bayesian CNN pre-trained to find morphological features on over $100$\,\text{M} responses by GZ (\citealp{Masters_2019}) volunteers answering a diverse set of classification tasks on over $800$\,\text{k} galaxies from the \textit{Sloan} Digital Sky Survey (SDSS), \HST (HST), Hyper Suprime-Cam (HSC), and the Dark Energy Camera Legacy Survey (DECaLS) imaging data  (\citealp{walmsley2024scaling}). 

\texttt{Zoobot} can be trained either from scratch or fine-tuned, a method in which a pre-trained model is adapted to solve a specific task (\citealp{Walmsley_2022}). The latter approach is the most common use of \texttt{Zoobot}, since it takes advantage of the learned representation of galaxy morphology. In this application to \Euclid ERO data, we use both approaches. In this work, we take the \texttt{EfficientNetB0} (\citealt{pmlr-v97-tan19a}) architecture variant of \texttt{Zoobot}, pre-trained on $100$\,\text{M} GZ classifications, and fine-tune it to classify strong lenses in the ERO data set.  Here, we fine-tune \texttt{Zoobot} by removing the `head' of the \texttt{EfficientNetB0} model and fixing the weights of the remaining layers (i.e. these layers remain `frozen'). We add a new head to the model with outputs appropriate to the new problem of classifying strong lenses; specifically, the binary classification task where the {model\textquotesingle}s  output is a predicted probability that a given input belongs to the positive class (i.e. class `lens'). The new head is trained to predict outputs for this new problem given the frozen representation and new labels, allowing the new head to benefit from previously learned representations. The new head is composed of two dense layers. The first dense layer has $128$ units with dropout probability $p=0.2$, and a rectified linear unit (ReLU) activation function. The second dense layer has two units with a softmax activation function.

Fine-tuning a classifier still requires some training examples, though less than training a classifier from scratch (\citealp{You2020CoTuningFT}). A great challenge in astronomy, specifically the research area of strong gravitational lensing, is the lack of sufficiently large labelled data sets to train supervised deep learning models. Similarly to previous studies (e.g \citealp[]{lanusse2018cmu,Huertas_Company_2020}), the model we describe in this section was trained using simulated images with known labels described in the next section.

\subsubsection{Simulated lenses for \texttt{Zoobot}}

The data used to train \texttt{Zoobot} for this application are simulated \Euclid images from the VIS instrument, created using the Python package \texttt{Lenzer}, specifically designed for this work.\footnote{\url{https://github.com/RubyPC/Lenzer}}
To create the simulations, \texttt{Lenzer} requires real galaxy images for the source galaxy to produce the simulated lensed light. These need to have at least the same spatial resolution and depth at approximately the same wavelength range as \IE-band (\citealp{Cropper_2016}) in order to simulate \Euclid-like data. These requirements are fulfilled with the Cosmic Evolution Survey (COSMOS; \citealp{2007ApJS..172....1S}) with the Advanced Camera for Surveys (ACS) Wide Field Channel of HST 
in the $F814W$ filter, with an angular resolution of 
\ang{;;0.09},
and a pixel scale of 
\mbox{\ang{;;0.03}}.
The depth and resolution are better than those estimated for \Euclid: $24.5$ mag at $10\sigma$ for \Euclid, as opposed to $27.2$ mag at $5\sigma$ for HST (\citealp{euclidcollaboration2024euclid});  however, the wavelength range of the \IE-band includes the $F814W$ band of HST, so we apply a magnitude restriction of $m_{F814W}<23.5$ with a redshift range of $2.5<z<5.0$, yielding a selection of $\num{36606}$ galaxies. We smooth the selected galaxy images with a kernel given by the difference in HST ACS and VIS point spread function and resample to give a \IE-band pixel scale of 
\ang{;;0.1}.
To simulate the lensed light, \texttt{Lenzer} requires the \texttt{Lenstronomy} Python package (\citealp{Birrer_2018}), \texttt{Skypy} (\citealp{amara2021skypy}) and \texttt{Speclite} which is part of the \texttt{Astropy} package (\citealp{astropy22}). To obtain a list of light profiles for \texttt{Lenstronomy}, we draw a random rotation angle for the galaxy images in our selection and transform the rotation angle and axis ratio into complex ellipticity moduli for the light profile. We use an elliptical Sérsic light profile for our lens simulations. Redshifts and magnitudes for the lens galaxies are sampled from a distribution of \textit{Sloan} Lens ACS Survey (SLACS) lens data (\citealp[]{2008ApJ...682..964B,Shu_2017}). From these sampled magnitudes, the velocity dispersions of the lens galaxies, $\sigma_v$, are calculated using the Faber--Jackson relationship (\citealp{faber1976velocity}), and the Einstein radius of each lensed image, $\theta_{\sfont{E}}$, is calculated using
\begin{equation}\label{eq:thetaE}
\theta_{\sfont{E}}=4\pi \left(\frac{\sigma_v}{c}\right)^2\frac{D_\sfont{LS}}{D_\sfont{S}},
\end{equation}
where $D_\sfont{LS}$ and $D_\sfont{S}$ represent the lens-source and the observer-source angular diameter distances, respectively. The remaining parameters of the lensed images are randomly sampled using a uniform distribution. Finally, Gaussian noise is added in each pixel to the individual simulated lensed images to reproduce the depth of the \IE-band. 
The simulated lens images are added to sources extracted from two other ERO fields, NGC6397 and NGC6822, with magnitudes \IE$ < 23$. These two ERO fields do not overlap the Perseus cluster. A total of $\num{50000}$ simulated lensed images were produced. The training set is also composed of $\num{50000}$ negatives from real \Euclid data. The negatives are sources extracted from the two ERO fields mentioned above, which were not used in creating the simulated lensed images.

Diversity of gravitational lensing systems is important to consider, and simulations of gravitational lenses using different methods can provide a range of different lensing configurations. The second data set we use to train from scratch and fine-tune \texttt{Zoobot} are lens simulations with known labels, created in the same manner as above. However, the simulated lensed images added to the source galaxies are produced in a different way. For a more detailed description of the simulation set, see Sect.\,3 of \cite{SGL_CNN_app_2019c}.

\subsubsection{Experiments and training procedure}

We thus trained the \texttt{EfficientNetB0} model with four different approaches and present a comparison of results in Sect.\,\ref{sec: Results}. In two of the approaches, we trained the \texttt{EfficientNetB0} model from scratch on two sets of simulated gravitational lenses, \texttt{Lenzer} and those described in \cite{SGL_CNN_app_2019c}. We will hereafter denote these networks with \texttt{E1} and \texttt{E2}, respectively.  For the remaining two approaches, we took the pre-trained network on the GZ classifications and fine-tune it with the same two simulation sets as before. We denote the fine-tuned networks as \texttt{Zoobot1} and \texttt{Zoobot2}, respectively. 

These four networks were trained in the same fashion. The images were resized to $300\times300$ pixels and normalised to have pixel values between $0$ and $1$. We trained each network up to $100$ epochs with train/validation/test sets split sizes of $80\%/10\%/10\%$, respectively. The test sets we describe here are not used for training, and are only used to evaluate the performance of each network on the two simulation sets. We used the 
binary-cross entropy
loss and the 
adaptive moment estimation
optimiser (\texttt{Adam}; \citealp{kingma2017adam}) with a learning rate of $10^{-4}$. \texttt{Adam} introduces an adaptive learning rate over standard optimisation methods such as 
stochastic gradient descent
(SGD). We applied a softmax activation to each {network\textquotesingle}s output to obtain a predicted probability of the positive class (class `lens'). We set a probability threshold of $p_{\sfont{THRESH}}=0.5$ for all networks for a given input to belong to the positive class, such that an image will be classified as a lens if the prediction of the network for the positive class is above $p_{\sfont{THRESH}}$. This is set by default in the \texttt{Zoobot} documentation but can be adapted (\citealp{Walmsley_2022}), although we kept the default probability threshold for this work. For each network, we saved the best weights which correspond to the epoch with the minimum validation loss, and we used early stopping to end training if the validation loss ceased to converge, thereby ending further training after epoch 92 for \texttt{Zoobot1}. The remaining three networks were trained for $100$ epochs.

\subsection{Naberrie and Mask R-CNNs}\label{ssec:Wilde}
The five networks discussed in this section are \texttt{Naberrie} (\textcolor{blue}{Wilde et al.} in prep), based on a U-Net architecture (\citealp{ronneberger2015unet}), and four Mask R-CNNs, or Mask Region-based CNNs, (\citealp{8237584}), namely \texttt{MRC-95}, \texttt{MRC-99}, \texttt{MRC-995}, and \texttt{MRC-3}. U-Net is a CNN originally designed for image segmentation (\citealp{ronneberger2015unet}), consisting of a contracting path and expansive path which are 
approximately 
symmetrical, giving the model its U-shape. The difference between U-Net and a generic CNN is the use of up-sampling whereby the pooling layers of a generic CNN are replaced by up-convolution layers, thus increasing the resolution of the output. Similar to U-Net, the Mask R-CNNs are designed for image segmentation but they also combine this with object detection. The Mask R-CNNs are able to perform pixel-wise segmentation using the `Mask' head which generates segmentation masks for each object detected (\citealp{8237584}).

\begin{figure}[t]%
    \centering
    {{\includegraphics[width=9cm]{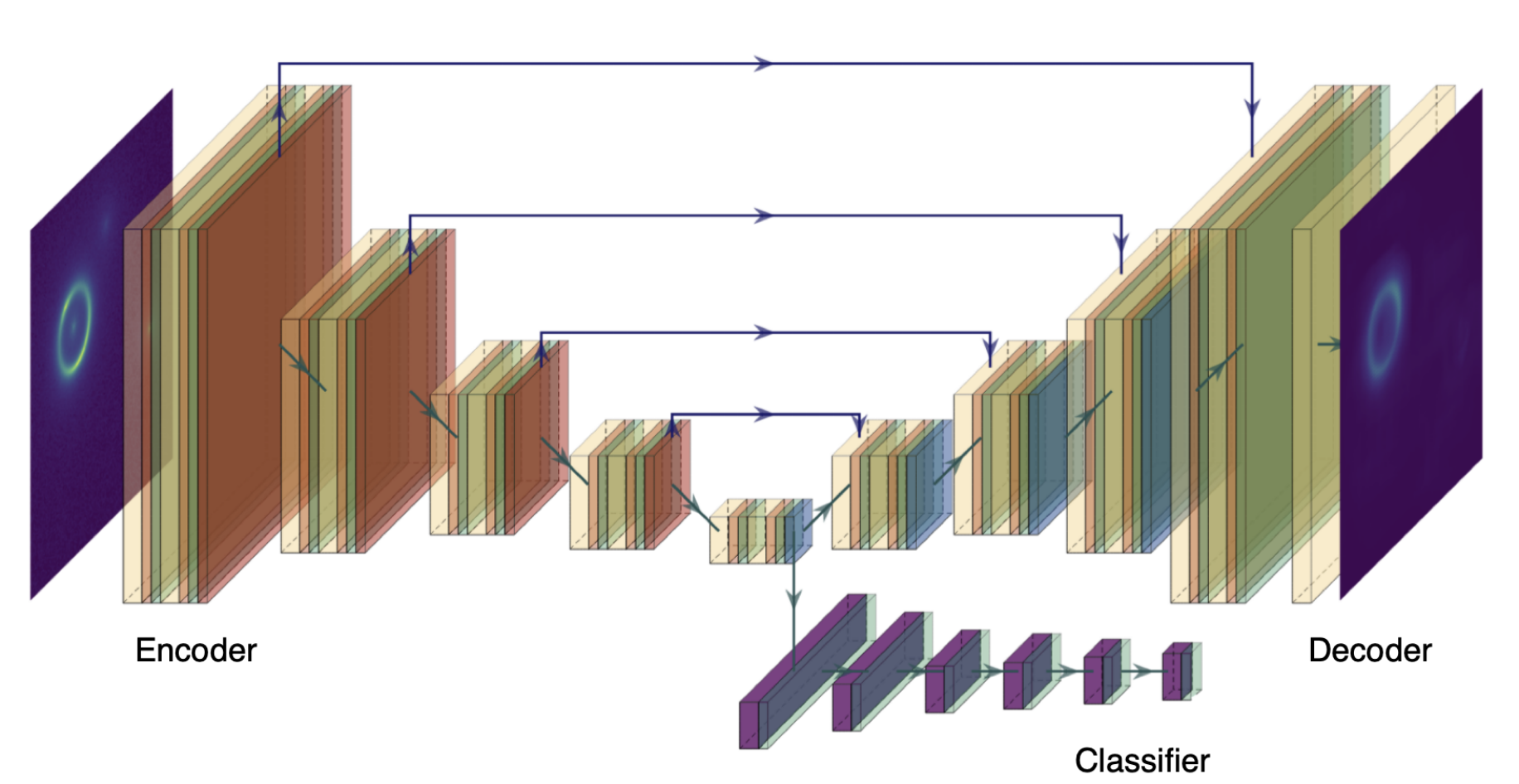} }}
    \caption{{The layers that make up the U-Net \texttt{Naberrie} including an encoder, a decoder, and a classifier. The arrows represent skip connections which concatenate extracted features from the encoder section directly to the decoder section. }}
    \label{fig:Naberrie}
\end{figure}

The U-Net \texttt{Naberrie} is made up of three separate parts: the encoder, the decoder, and the classifier. Figure\,\ref{fig:Naberrie} shows its architecture. \texttt{Naberrie} takes an \IE-band image of $200 \times 200$ pixels, corresponding to $20^{\prime\prime}\times20^{\prime\prime}$ with an \IE-band pixel scale of $\ang{;;0.1}$. The encoder section is made of several convolutional layers with max-pooling and dropout applied. This produces a bottleneck that is a dimension-reduced representation of the input data. The decoder section is trained to take the bottleneck data as input and to produce an image as an output. The target for the decoder is to produce an image showing only the light from the lensed galaxy in the input image. The classifier also takes the bottleneck data as input; the goal of the classifier is to classify the input image as either a lens (1) or a non-lens (0). The classifier is made up of fully-connected layers and 
ReLU
activation functions.

The Mask R-CNNs all have the same \texttt{ResNet-50} architecture (\textcolor{blue}{\citealp{he_2016}}) where the default final layer for each model is replaced with a sigmoid activation to output a probability of an image belonging to class lens (1) or non-lens (0), with  the rest of the model weights frozen. The main difference between the variations of the Mask R-CNNs is the threshold of the masking ($0.95$, $0.99$, and $0.995$, respectively);  \texttt{MRC-3} also had a mask threshold of $0.995$. For more information, we refer the reader to \cite{wilde2023applications}.

The simulated data used to train the models \texttt{Naberrie}, \texttt{MRC-95}, \texttt{MRC-99}, \texttt{MRC-995}, and \texttt{MRC-3} were created using \texttt{Lenstronomy} (\textcolor{blue}{\citealp{Birrer_2018}}) with lens parameters taken from \cite{SGL_CNN_app_2019c}. The aims of these simulations were to train models that could be more interpretable than the standard CNN by providing a location for lensing features through building interpretability into the model, since the Mask R-CNNs use object detection methods, rather than interpretability techniques being used post-training (\textcolor{blue}{\citealp{jacobs2022exploring, wilde2022detecting}}).  The redshifts and magnitudes were sampled from the second strong gravitational lens finding challenge training data set (\citealp{SGL_CNN_app_2019c}), which involve 
similar 
but not identical
lens parameters as those used in the simulations in Sect.\,\ref{ssec:Zoobot} for networks \texttt{Zoobot2} and \texttt{E2}. From this, the velocity dispersion was determined (\textcolor{blue}{\citealp{faber1976velocity}}) and the Einstein radius calculated. Other galaxy parameters were sampled from a uniform sampling of Sloan Lens ACS survey data (SLACS \textcolor{blue}{\citealp{2008ApJ...682..964B,auger2009sloan,newton2011sloan,Shu_2017,denzel2021new}}). The training set contained a total of $\num{10000}$ cutouts. \texttt{Naberrie} requires a target of the lens light for the decoder, and the Mask region-based CNNs (Mask R-CNNs) require a mask for each galaxy. The full details of this, and the validation and test sets used, can be found in \cite{wilde2023applications}.

\texttt{Naberrie} was trained in two stages both using $\num{10000}$ \IE-band $20^{\prime\prime}\times20^{\prime\prime}$ cutouts for $50$ epochs with a learning rate of $3\times 10^{-4}$, and a batch size of $10$. In the first stage, we trained the encoder and the decoder. Once these had been trained, the best weights from the epoch with the lowest validation loss are saved,  and loaded into \texttt{Naberrie}. In the second stage, the training is done after freezing the encoder weights to the best weights from the previous stage, and replacing the decoder by the classifier. Finally, 
all three sections of \texttt{Naberrie} were assembled to form the final trained model. 
The Mask R-CNNs were trained using $\num{10000}$ \IE-band $20^{\prime\prime}\times20^{\prime\prime}$ cutouts with a learning rate of $3\times 10^{-4}$, and a batch size of $10$, for $50$ epochs (for \texttt{MRC-95}, \texttt{MRC-95}, \texttt{MRC-995}) and $100$ epochs (for \text{MRC-3}). The Mask R-CNN outputs a probability for each class, a mask for the detection, and a bounding box for the detection. The classifiers described in this section were trained with the binary-cross entropy loss optimised by the \texttt{Adam} optimisation method (\citealp{kingma2017adam}).

\subsection{NG (Napoli-Groningen) and ResNet-18}\label{ssec: NG}
The Napoli-Groningen network, \texttt{NG}, is a CNN with an architecture based on the model described in \cite{Petrillo_2}, employing a \texttt{ResNet-18} architecture \citep{he_2016}. The network takes an input in the form of a $200\times 200$ pixel image, corresponding to $20 \arcsecond \times 20 \arcsecond$ for an \IE-band pixel scale of $\ang{;;0.1}$, and outputs a number between $0$ and $1$, which represents the probability of the image being a candidate lens. The training data consists of approximately \num{6000} lenses, and \num{12000} non-lenses. The lenses are created from $r$-band KiDS luminous red galaxies (LRGs) by super-imposing $10^{6}$ simulated lensed objects. The simulated lensing features are mainly rings, arcs, and quads. The gravitational lens mass distribution adopted for the simulations is assumed to be that of a Singular Isothermal Ellipsoid (SIE), perturbed by additional Gaussian random field fluctuations and external shear. An elliptical Sérsic brightness profile is used to represent the lensed sources, to which several small internal structures were added (e.g. star-formation regions), described by circular Sérsic profiles. The non-lenses in the training set are galaxies from KiDS composed of normal LRGs without lensing features, randomly selected galaxies from the survey with $\textit{r}\,<21$ magnitude, mergers, ring galaxies, and a sample of galaxies that were visually classified as spirals from a GZ project (\citealp{Willett2013}). \texttt{NG} was trained with the binary-cross entropy loss optimised by the \texttt{Adam} optimisation method (\citealp{kingma2017adam}) with a learning rate of $4\times 10^{-4}$ for 100 epochs. For further details see \citet[]{Petrillo_2,Petrillo_3}.

Additionally, the same \texttt{ResNet} architecture was trained on a different set of lens simulations described in Sect.\,3 of \cite{SGL_CNN_app_2019c}, namely \texttt{ResNet-18}, with the final three layers replaced by fully-connected dense layers followed by a sigmoid activation. We trained the network on $\num{80000}$ simulated gravitational lenses, and corresponding non-lenses, of size $5 \arcsecond \times 5 \arcsecond$, in the same manner for \texttt{NG} as described above.

\subsection{4-layer CNN}\label{ssec:4layer}

\begin{figure*}[t]%
    \centering
    \includegraphics[width=18cm]{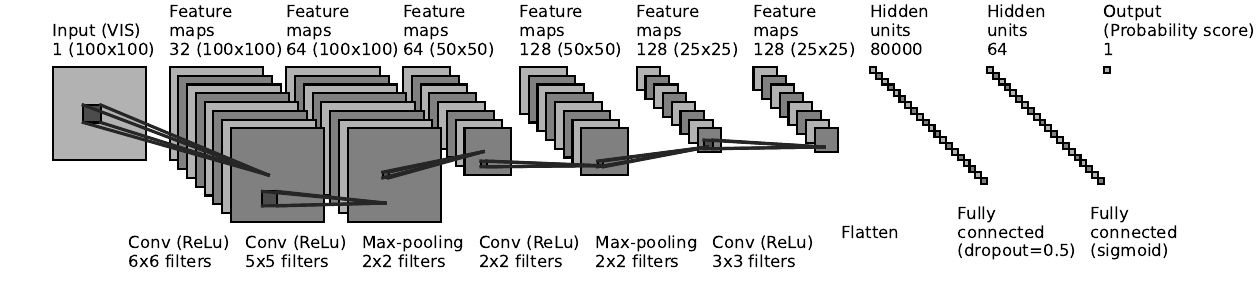}
    \caption{Scheme of the architecture of the \texttt{4-layer CNN}. The \texttt{4-layer CNN} takes a $100\times 100$ pixel \IE-band image as input and extracts features through blocks of convolution layers and max-pooling layers, with ReLU activation functions. Convolution layer sizes are stated in Sect.\,\ref{ssec:4layer}. The output of the final convolution layer is flattened and is input to a fully-connected layer with dropout probability $p=0.5$. The final fully-connected layer uses a sigmoid activation to produce a probability score as output, with 1 belonging to class lens.}
    \label{fig:4-layer_CNN}
\end{figure*}

We test the performance of a shallow \texttt{4-layer CNN} in identifying strong lens candidates in the ERO data of the Perseus cluster. The CNN architecture used, shown in Fig.\,\ref{fig:4-layer_CNN}, is an adaptation of the network developed for the morphological classification of galaxies in \cite{DominguezSanchez2018}, and has already been tested with \Euclid-like simulations \citep{manjon_thesis_2021}. The network has four convolutional layers of different spacings (6$\times$6, 5$\times$5, 2$\times$2, and 3$\times$3,
respectively), and two fully-connected layers. ReLU activation functions are applied after
every convolutional layer, and a $2\times2$ max-pooling is applied after the second and third convolutional layers. 

The training of the network was carried out following a supervised learning approach with labelled \IE-band images. The training data set for the network was strong lens simulations created by Metcalf et al. (\citeyear{SGL_CNN_app_2019c}, 2024, in prep.), which include companions that are also lensed. It consists of $\num{60000}$ images split into $90\%$ for training and $10\%$ for validation, with a balanced distribution of lenses and non-lenses. We investigated training under different conditions (simulation data sets used, training set sizes, data pre-processing, small changes in the architecture and learning parameters) in order to achieve both the lowest loss and false positive rate, whilst maintaining a high true positive recovery. The best network, however, yielded many false positives when tested on the Perseus cluster, which will be described in detail in Sect.\,\ref{sec: Results}. The network reads \IE-band images of $100\times100$ pixels, corresponding to $10 \arcsecond \times 10 \arcsecond$ in \IE-band, in which the fluxes are normalised to the maximum value. In the learning process, we used a binary cross-entropy loss function optimised by the \texttt{Adam} stochastic optimisation method (\citealp{kingma2017adam}), with a learning rate of $10^{-3}$. The model was trained for $20$ epochs, using a batch size of $24$. In order to prevent over-fitting, several data augmentation techniques were performed during training, allowing the images to be zoomed in and out ($0.75$ to $1.3$ times the original size), rotated, and flipped and shifted both vertically and horizontally. Finally, in the last fully-connected layer, a sigmoid activation function is used to turn the output scores into probabilities of being a lens distributed between $0$ and $1$, with $1$ denoting the class `lens'. 

\subsection{Denselens and U-Denselens}\label{ssec:Denselens}
\texttt{Denselens} (\citealp{nagam}) is an ensemble of classification and regression networks to classify and rank strong lenses. \texttt{Denselens} is constructed with a densely connected convolutional network (DenseNets, \citealp{huang2017densely}) architecture as a backbone in which the layers are connected in a feed-forward manner. Here, we describe the \texttt{Denselens} network and an adaptation of this network which we have called \texttt{U-Denselens}, where an additional U-Net has been added to the original \texttt{Denselens} architecture. We take the frozen weights of the pre-trained \texttt{Denselens} network for this application.

The \texttt{Denselens} network includes an ensemble of four classification prediction networks in which each network predicts an output value in the range $[0, 1]$. The mean of the output predictions from the four networks, $P_\textrm{mean}$, are calculated and the candidates whose predictions fall above a threshold ($p_{\sfont{THRESH}}$) are classified as strong lens candidates, while the rest are classified as non-lenses. We set the $p_{\sfont{THRESH}}$ value to be $0.23$ for this application to find strong lenses, since this value of $p_{\sfont{THRESH}}$ maintained a true positive recovery of $50\%$. The candidates that are classified as lenses are then passed to the second set of ensemble networks comprising four CNNs, trained on the Information Content (IC) value of mock simulated lenses. The IC values used for training lenses is calculated based on Eq.\,(\ref{eq:IC}): 
\begin{equation}\label{eq:IC}
    \textrm{IC} =  \left [\frac{ A_{\textrm{src,}2\sigma} }{A_\sfont{PSF}} \right] R, 
\end{equation}
where 
$A_{\textrm{src,}2\sigma}$ 
is the total area of the lensed images above a brightness threshold in terms of noise level of $2\,\sigma$. The area of the PSF ($A_\sfont{PSF}$) is defined as the square of Full Width Half Maximum (FWHM) and $R=(\theta_{\sfont{E}}/R_\textrm{eff})$ is the ratio of the Einstein radius ($\theta_\sfont{E}$) to the effective source radius ($R_\textrm{eff}$). In practice, \text{IC} can be used as an effective metric to rank-order lenses that are easy to recognise for a human or CNN classifier. 

In addition to this, in our second network, \texttt{U-Denselens} (\citealp{nagam2}), we have included a U-Net (\citealp{ronneberger2015unet}) at the end of the \texttt{Denselens} pipeline, to further improve the accuracy in detecting strong lenses and reducing the false positive rate. The U-Net network is trained by designating all source pixels to a value of $1$ and all other pixels to $0$. The output of the U-Net corresponds to a map of $101 \times 101$ pixels. Each pixel in the output map varies in the range $[0,1]$ since a sigmoid function has been used as an activation function in the final layer. Using the information from the source pixels, we define a metric, $n_\textrm{s}$, which is calculated by the number of source pixels above a segmentation threshold, $S_\sfont{THRESH}$. This segmentation threshold represents the pixels that are considered to belong to the source by U-Net. Setting the threshold too low would result in selecting pixels for which the U-Net lacks confidence. Conversely, a high threshold could lead to a multitude of candidates with minimal pixels in the segmentation output. We have set the $S_\sfont{THRESH}$ for \texttt{Denselens} and \texttt{U-Denselens} to a small value of $0.01$, and we set $p_{\sfont{THRESH}}=0.18$ for \texttt{U-Denselens} for this application to the Perseus cluster. For more information, we refer the reader to \cite{nagam2}.
The candidates that have passed the $p_{\sfont{THRESH}}$ and have $n_\textrm{s}$ $>0$ are selected as lens candidates. The candidates that were rejected with $n_\textrm{s} = 0$, were only retained if they have $\text{IC}\geq88$. This step of retaining based on the \text{IC} is done to reduce the rejection bias of the segmentation network since the network is not trained with all possible combinations of extremely diverse lens samples present. 

KiDS data were used to create the simulated data to train \texttt{Denselens} and \texttt{U-Denselens}. The training data set comprises $\num{10000}$ samples, balanced between \num{5000} mock lenses and \num{5000} non-lenses. The source galaxies are modelled by sampling parameters from a Sérsic profile, such as the effective radius $R_{\textrm{eff}}$, and the Sérsic index $n$. The lens galaxies are modelled with an SIE mass distribution. For more information on the lensing simulations used to train \texttt{Denselens}, see \cite{nagam,nagam2}. The mock lenses are painted over the roughly \mbox{$5000$} selected LRGs from the KiDS data set.  The negatives comprise \mbox{$5000$} galaxies, which include random galaxies, sources that have been wrongly identified as mock lenses in previous tests, and also visually classified spiral galaxies from a GZ project. Each simulated  mock lens and  non-lens corresponds to an area of $20^{\prime\prime}\times20^{\prime\prime}$. The training strategy of \texttt{Denselens} and \texttt{U-Denselens}, and more information regarding the simulations used for training, and validation and test sets, are explained in detail in \cite{nagam,nagam2}, respectively.

\subsection{LensCLR}\label{ssec: lensclr}

Here, we explore the application of semi-supervised learning for detecting lens candidates, as it offers a compelling advantage over fully-supervised learning by leveraging both labelled and unlabelled data to enhance model performance \citep[e.g.,][]{2022ApJ...932..107S,2023RASTI...2..441H}.
This approach addresses the challenge of acquiring large labelled data sets, which are often costly and time-consuming to obtain. 
Semi-supervised learning involves two training steps.
First, the network is trained using unlabelled images to learn the underlying data patterns and structures in a self-supervised manner via contrastive learning. 
The aim of this process is to cluster similar images together in the latent space while pushing dis-similar images apart \citep{2020arXiv200205709C}.
Then, the pre-trained network is fine-tuned for the classification task using a small number of labelled images.
This strategy not only reduces labelling costs but also improves generalisation and robustness, since the models can capture a wider variety of patterns from diverse unlabelled data \citep{2020arXiv200610029C,2020arXiv200304297C}.

The details of our semi-supervised network architecture, hereafter \texttt{LensCLR}, will be presented in a forthcoming paper (Andika et al. 2024, in prep.). 
Our model follows the simple framework for contrastive learning of visual representations proposed by \cite{2020arXiv200205709C}. 
\texttt{LensCLR} comprises three main components: augmentation, backbone, and projector modules.
For the backbone, we use \texttt{EfficientNetV2B0} to transform each input image into a lower-dimensional representation, yielding a $\num{1280}$-dimensional vector as output \citep{2021arXiv210400298T,Andika_2023}. 
Following that, we attach two dense layers, each with $128$ neurons, as the projector; collectively, the backbone and the projector form the encoder. 
Finally, the encoder learns meaningful representations by employing a contrastive loss function, associating augmented views of the same image as similar and views of different images as dissimilar \citep{2018arXiv180703748V}.

The first phase of the training, also called pre-training or the pretext task of \texttt{LensCLR}, via contrastive learning, is done by using an unlabelled data set containing around $\num{500000}$ sources in the Perseus cluster.
This data set, called \text{DS}1, includes real galaxies, stars, quasars, photometric artefacts, and other features observed in \IE images.
In the training loop, batches of 512 images of $150 \times 150$ pixels are randomly selected from \text{DS}1 as inputs, and their fluxes are scaled to be between $0$ and $1$ using min-max normalisation.
These images are then passed through the augmentation module, which transforms the inputs with random flips, $\pm \pi/2$ rotations, and $5$-pixel shifts, as well as a centre-crop to trim them to $96\times96$ pixels.

Note that augmenting sample $\vec{x}_q$ will produce a pair of views marked as positive ($\vec{x}_q$, $\vec{x}_{k^+}$) if they originate from transformations of the same source, i.e. there is a (positive) similarity since the pair of views originate from the same source with the difference being that one is augmented, and negative ($\vec{x}_q$, $\vec{x}_{k^-}$) otherwise, since the pair of views originate from different sources and are thus dissimilar \citep[e.g.][]{2021ApJ...911L..33H}.
Next, these images are processed by the encoder, where the output is projected into a $128$-dimensional vector: $\vec{z}=\mathrm{encoder(\vec{x})}$.
Model optimisation is then performed, starting with an initial learning rate of $10^{-3}$, by minimising the contrastive loss given by:
\begin{equation}
\label{eq:contrastive_loss}
L_{q, k^+, \{k^-\}}= -\ln  \left\{
	 \frac{
	 	\exp[ \mathrm{sim}(\vec{z}_q , \vec{z}_{k^+})]
	 }
	 {
	 	\exp[ \mathrm{sim}(\vec{z}_q , \vec{z}_{k^+})] + \sum_{k^-} \exp[ \mathrm{sim} (\vec{z}_q , \vec{z}_{k^-})]
	 } 
\right\}.
\end{equation}
Note that $\text{sim}(\vec{a}, \vec{b}) = \vec{a} \cdot \vec{b} / (\tau\ | \vec{a} | \ | \vec{b} |)$ represents the cosine similarity between vectors $\vec{a}$ and $\vec{b}$, scaled by a configurable `temperature' parameter $\tau$. 
This loss function, also known as information noise contrastive estimation \citep[InfoNCE;][]{2018arXiv180703748V}, is optimised to ensure that similar pairs (positive pairs) exhibit high similarity, whereas dissimilar pairs (negative pairs) display low similarity.

In the second phase of training, we fine-tune \texttt{LensCLR} using a supervised learning approach with labelled images. 
The labelled data set, referred to as \text{DS}2, is derived from the strong lens simulation created by Metcalf et al. (\citeyear{SGL_CNN_app_2019c}, 2024, in prep.), containing \num{50000} lens and \num{50000} non-lens systems, identical to other training sets used in this work. For a more detailed description of the simulation set, see Sect.\,3 of \cite{SGL_CNN_app_2019c}.
We split \text{DS}2 into training, validation, and test data sets with a ratio of $70\%$:$20\%$:$10\%$.
After that, a modification needs to be applied to the encoder by removing the projector and replacing it with a single dense layer consisting of one neuron with a sigmoid activation function.
This single dense layer will later become the classification layer. 
During fine-tuning, all weights in the pre-trained encoder are frozen, and only the classification layer is trained. 
The objective is to differentiate between lenses and non-lenses using a binary cross-entropy loss function optimised with the \texttt{Adam} optimiser (\citealp{kingma2017adam}), and initial learning rate of $10^{-4}$.

During the training loop, we implement a scheduler that reduces the learning rate by a factor of 0.2 if model performance does not improve over five consecutive epochs \citep[e.g.,][]{2023ApJ...943..150A}. 
Additionally, early stopping is used if the loss does not decrease by at least $10^{-4}$ for 10 consecutive epochs. 
After 200 epochs, our model converges, and the optimised parameters are saved.

\subsection{VGG, GNet, and ResNext}\label{ssec:Leuzzi}
In \cite{Leuzzi-TBD}, we applied three network architectures to find lenses in simulated \Euclid-like images: a \texttt{VGG}-like network, a \texttt{GNet} which includes an inception module, and a residual network \texttt{ResNext}. We now apply all the architectures to the ERO Perseus cluster data set. 

While we refer the reader to \cite{Leuzzi-TBD} for a detailed description of the architectures, we summarise their main characteristics here.  The \texttt{VGG}-like network (inspired by \citealp{simonyan_2015}) has ten convolutional layers alternating with five max-pooling layers, followed by two fully-connected layers. The main building block of \texttt{GNet} is the inception module introduced by \citet[]{szegedy_2015,szegedy_2016}. The inception module is implemented as a series of convolutions of different sizes ($1\times1$, $3\times3$, $5\times5$) run in parallel on the same image to extract features on different scales simultaneously. Our implementation of this architecture starts with two convolutional layers alternating with two max-pooling layers. After this, there is a sequence of seven inception modules, the fifth of which is connected to an additional classifier. When computing the loss function during training, the input of the fifth and of the final output layers are combined, with different weights ($0.3$ and $1.0$, respectively). The \texttt{ResNext} is based on the concept of residual learning introduced by \cite{he_2016} and \citet{Xie2016AggregatedRT}, and is implemented in such a way that every building block of the architecture does not have to infer the full mapping between the input and output, but only the residual mapping. This is implemented through shortcut connections (\citealp{he_2016}). Our version of this architecture comprises two convolutional layers and four residual blocks alternating with two max-pooling layers. The final layer of all networks uses the sigmoid function to provide a value in the range $0$ to $1$, used for the classification of the image as a lens or non-lens as a probability score, given the desired threshold (we set $p_{\sfont{THRESH}} =  0.5$).

We train the networks with $\num{100000}$ \Euclid-like mock images, created by Metcalf et al. (\citeyear{SGL_CNN_app_2019c}, 2024, in prep.). The simulations are based on the halo catalogues provided by the Flagship simulation \citep{EuclidSkyFlagship}.  All networks are trained for $100$ epochs, with an initial learning rate of $10^{-4}$, adjusted during training to improve convergence, depending on the trend of the validation loss function over time. We use the \texttt{Adam} optimiser (\citealp{kingma2017adam}) and the binary cross-entropy loss. The pre-processing of the images and training procedure are further detailed in Sect.\, 4 of \cite{Leuzzi-TBD}. 

\subsection{Consensus-Based Ensemble Network (CBEN)}\label{ssec:cben}

The Consensus-Based Ensemble Network (\texttt{CBEN}) leverages the complementary strengths of seven CNNs: \texttt{ResNet-18} (\citealp{he_2016}); \texttt{EfficientNetB0} (\citealp{pmlr-v97-tan19a}); \texttt{EfficientNetV2} (\citealp{2021arXiv210400298T}); \texttt{GhostNet} (\citealp{2021arXiv210400298T}); \texttt{MobileNetV3} (\citealp{howard2019searching}); \texttt{RegNet} (\citealp{radosavovic2020designing}); and Squeeze-and-Excitation \texttt{ResNet} (\texttt{SE-ResNet}, \citealp{hu2018squeeze}). Each of these networks has been independently trained to classify lenses. By incorporating diverse architectures within the ensemble, \texttt{CBEN} benefits from the varied feature extraction capabilities and decision-making heuristics of the individual models. 

The standout feature of \texttt{CBEN} is its decision-making process, which relies on a unanimous voting mechanism, where a data sample is classified as belonging to the positive class only if all seven models independently agree on this prediction, i.e. the prediction, $p>\beta$, for $0<\beta<1$. In that case, the model score is taken as the minimum of all model scores. Conversely, if there is any disagreement among the models, the sample is classified as belonging to the negative class, with the model score taken as the maximum of all model scores. This criterion ensures a high level of confidence and reduces the likelihood of false positives, which is critical in lens identification where the cost of a false positive is high. We investigated the performance of \texttt{CBEN} with $\beta=0.3$ and $\beta=0.0001$, hereafter \texttt{CBEN-0.3} and \texttt{CBEN-0.0001}, respectively.

All networks were trained on mock images resembling those of \Euclid \IE-band, simulated in accordance with the methods described in Metcalf et al. (\citeyear{SGL_CNN_app_2019c}, 2024, in prep.), incorporating enhanced noise and PSF modelling. The training data set consists of $\num{80000}$ images, with $\num{10000}$ samples used for validation and another $\num{10000}$ for testing. The images were pre-processed by cropping them to dimensions of $96\times96$ pixels and normalising their pixel values to a range of $[0,1]$. During training, we employed the \texttt{Adam} optimiser (\citealp{kingma2017adam}) with a learning rate of $10^{-4}$, and binary cross-entropy loss. After 100 epochs, the models with the lowest validation error were saved and tested on the common test set.

\section{Metrics used for analysis}\label{sec:metrics}
The performance of a network can be assessed in many ways, however we focus both on the number of true positives (\text{TPs}) recovered, the and false positives (FPs) found from the common test set of \num{12086} stamps from the Perseus cluster. We first define the true positive rate and false positive rate metrics (hereafter \text{TPR} and \text{FPR}, respectively), which can be used to visualise a network's performance using a Receiver Operating Characteristic curve (ROC). Then, using the ROC curve, we can calculate another metric known as the Area Under the ROC curve (AUROC). Here, we give an overview of these metrics which we use to evaluate the performance of the various networks, in addition to introducing the importance of interpreting the false positives.

\subsection{\text{TPR} and \text{FPR}}\label{ssec:tpr fpr}
In our context, a true positive refers to a candidate lens correctly identified as belonging to the positive class (i.e. class `lens'), and a false positive represents a non-lens falsely identified as belonging to the positive class. We use the following definitions of true positive rate (\text{TPR}, also known as 
recall
or 
sensitivity) and false positive rate (\text{FPR}, 
also known as 
contamination or specificity) at a given threshold:
\begin{equation}\label{eq: metrics}
\text{TPR}=\frac{N_{\sfont{TP}}}{N_{\sfont{TP}}+N_{\sfont{FN}}},  
\quad 
\text{FPR}=\frac{N_{\sfont{FP}}}{N_{\sfont{FP}}+N_{\sfont{TN}}},
\end{equation}
where $N_\sfont{TP}$ is the number of true positive (\text{TP}) stamps that were correctly predicted, $N_\sfont{TN}$ is the number of true negative (\text{TN}) stamps that were correctly predicted, and $N_\sfont{FN}$ indicates the number of stamps that belong to the positive group but were falsely classified as negative (false negatives, or \text{FN}). Finally, $N_\sfont{FP}$ represents the number of stamps that belonged to the negative group but were false positives (\text{FP}). For our application to the $\num{12086}$ stamps from the Perseus cluster, \text{TPR} represents to proportion of grade $A$, grade $B$, and grade $C$ lens candidates correctly identified, and \text{FPR} represents the proportion of stamps incorrectly classified as a lens.  

\subsection{ROC}\label{ssec:roc metric}
We can visualise a network's classification performance at every possible threshold using a Receiver Operating Characteristic curve (ROC) which shows the \text{TPR} vs \text{FPR}, that were defined in the previous section. A binary classifier generally gives a probability of a candidate being a lens, $p_{\sfont{LENS}}$, in which case a threshold is set, namely $p_{\sfont{THRESH}}$, and everything with $p_{\sfont{LENS}}>p_{\sfont{THRESH}}$ is classified as a lens.
At $p_{\sfont{LENS}}=0$, all of the cases are classified as non-lenses and so \text{TPR} $ = $ \text{FPR} $= 0$, and at $p_{\sfont{LENS}}=1$, all of the cases are classified as lenses so \text{TPR} $=$ \text{FPR} $= 1$. These points are always added to the ROC curve. If the classifier made random guesses then the ratio of lenses to non-lenses would be the same as the ratio of the number of stamps classified as lenses to the number of stamps classified as non-lenses and so \text{TPR} $=$ \text{FPR}. For our application to the Perseus cluster, the best classifier will have a small \text{FPR} and a large \text{TPR}, and so the further away from this diagonal line it will be.

\subsection{AUROC score}\label{ssec:auc metric}

From the ROC curve described in the previous section, we can calculate the Area Under the ROC (AUROC) score, another metric which is widely used in other studies (\citealp{SGL_CNN_app_2019c}). A large AUROC score means that many true positives can be recovered at low false positive rates. A small AUROC score means that high false positive rates are required to achieve high true positive recovery. AUROC ranges between $0$ and $1$, where a value of AUROC close to $0.5$, and below, show a poor classification, with an AUROC score of $0$ meaning that the classifier is always wrong in its classification. An AUROC score of 1 represents a perfect classifier (for more information about ROC and AUROC metrics see \citealp{teimoorinia2016artificial} and references therein). Our best classifier will have a high AUROC score.

\subsection{Interpreting false positives}\label{ssec:fps interpretation}
In addition to the important metrics defined above, we will also give an interpretation of the false positives in the results. Comparing the false positive rate, and the number of false positives (or the contamination), of each network allows us to access the classification performance when the networks, previously trained on simulations, are applied to real data. The networks with a low false positive rate will have a low contamination (i.e. they will find few false positives). These networks will therefore have a lower contamination rate, defined by $N_\sfont{FP}/N_\sfont{TP}$, and will produce a more pure sample of lenses. The contamination rate is an important metric to consider, since networks with large contamination rates will find large numbers of false positives if applied to future \Euclid observations, beyond which a visual inspection is infeasible. 

Additionally, an investigation into the most frequent galaxy morphologies contaminating the classification result across all networks indicates which features the networks are responding to, and allows us to proceed by introducing these negatives in future training of the networks.

\section{Results}\label{sec: Results}
Here, we attempt a systematic comparison of network performance in preparation for CNN-based detection pipelines for the VIS images of the ERO data of the Perseus cluster using the metrics defined in Sect.\,\ref{sec:metrics} to compute the ROC of each network. We compute the AUROC score for each network and compare the \text{FPR} and \text{TPR} of the networks. Additionally, we compare the lens misclassifications which are common to the majority of our networks.

\subsection{ROC and AUROC}\label{ssec:rocauc}
 In total, $20$ networks were trained with their respective training sets and evaluated on the common test set of $\num{12086}$ real ERO stamps from the Perseus cluster that were visually inspected. Here, we characterise and compare the performance of the various networks in terms of the ROC and AUROC score as defined in the previous section.

\begin{figure*}[!htb]%
    \centering
    {{\includegraphics[width=18.35cm]{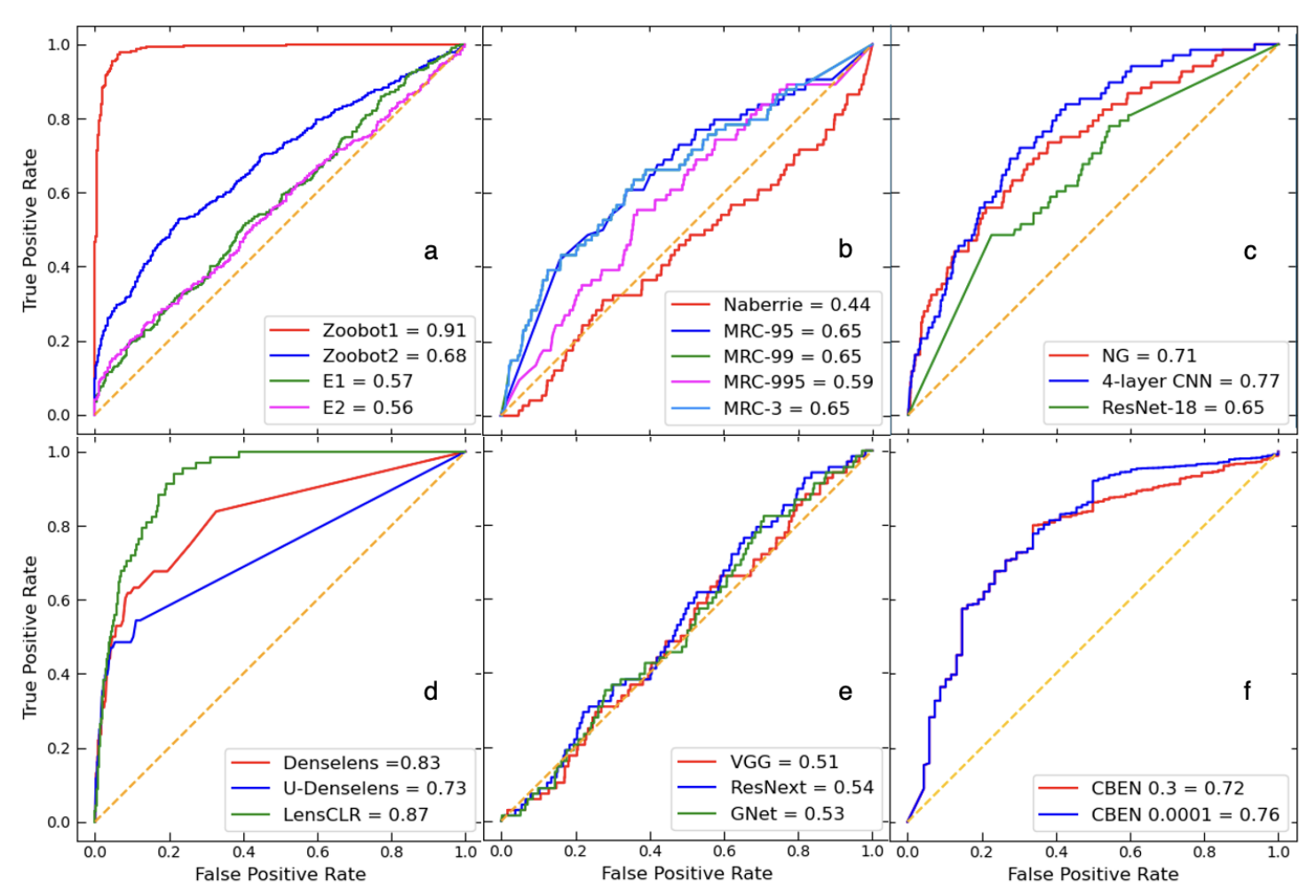} }}
    \caption{ROC curves for the networks described in Section \ref{sec: Methods}.  Performance of the \texttt{Zoobot} networks are shown in panel (a), \texttt{Naberrie} and \texttt{MRC} networks are shown in (b), \texttt{NG}, \texttt{4-layer CNN}, and \texttt{ResNet-18} are shown in (c), and \texttt{Denselens} and \texttt{LencCLR} are shown in (d). The performance of the networks described in Sect.\,\ref{ssec:Leuzzi} are shown in (e), and (f) shows the \texttt{CBEN} networks at two voting values $\beta=0.3$ and $\beta=0.0001$, respectively. The orange diagonal line marked represents the expected performance of an entirely random classifier. The values in the legend of each subplot represent the true positive recovery for each network. Higher performance is represented in general by curves approaching the top-left corner of the plots. The best performing gravitational lens classifiers are the \texttt{Zoobot1} and \texttt{LensCLR} networks.}
    \label{fig:ROCAUC}
\end{figure*}

Figure\,\ref{fig:ROCAUC} shows the ROC curves for all networks evaluated on the common test set. The random classifier (orange dotted line) serves as a reference. The ROC curves represent the proportion of the $68$ lens candidates from the expert visual inspection correctly identified as lens candidates against the proportion of non-lenses in the expert visual inspection sample falsely identified as lens candidates by each network. To quantify the performance of each network, we compute 
the AUROC score. A perfect binary classifier would have an AUROC score of $1.0$, meaning that the classifier can distinguish between all positive and negative class points.

Figure\,\ref{fig:ROCAUC} (a) shows the performance of the four \texttt{EfficientNetB0} networks, of which two are pre-trained on GZ classifications and fine-tuned on the separate simulation sets described in Sect.\,\ref{ssec:Zoobot}, namely \texttt{Zoobot1} and \texttt{Zoobot2}, shown by the red and blue lines, respectively. The remaining two networks are trained from scratch on the same two separate simulation sets, \texttt{E1} and \texttt{E2}, shown by the green and magenta lines, respectively. The fine-tuned \texttt{EfficientNetB0} networks, namely \texttt{Zoobot1} and \texttt{Zoobot2}, appear to be effective classifiers and are more confident in ranking a candidate lens higher than a non-lens than \texttt{E1} and \texttt{E2}. The best performing network out of the \texttt{EfficientNetB0} networks is \texttt{Zoobot1}, which achieves an AUROC score of $0.91$. The fact that \texttt{Zoobot1} is confident at identifying lens candidates is most likely attributed to its pre-training on $800$\,\text{k} galaxy images with over $100$\,\text{M} volunteer responses and fine-tuning on training data sets given by \texttt{Lenzer}. However, the fine-tuned \texttt{EfficientNetB0} network trained on a different set of simulations, \texttt{Zoobot2}, only achieves an AUROC score of $0.68$. This reinforces the fact that the training sample used to fine-tune the networks has a significant impact on the performance. This becomes evident when we compare the results of the \texttt{Zoobot} networks trained from scratch without the learned representation of galaxy morphology from GZ on the same simulation sets, namely \texttt{E1} and \texttt{E2}, which are worse at correctly classifying the lens candidates.
 
 Figure\,\ref{fig:ROCAUC} (b) shows the ROC curves for the networks described in Sect.\,\ref{ssec:Wilde}, which are trained on similar simulation sets to \texttt{Zoobot2} and \texttt{E2}. These networks do not appear to be strong classifiers. The best-performing model in Fig.\,\ref{fig:ROCAUC} (b) is \texttt{MRC-3} with an AUROC score of $0.66$. Although \texttt{MRC-3} achieves a relatively low AUROC score, it out-performs \texttt{Naberrie}, which achieves an AUROC score of $0.45$ and performs worse than a classifier that ranks lenses and non-lenses at random.  Thus, it is possible that the network architecture is an important factor in performance, in addition to the training sample, with {\texttt{MRC-3}\textquotesingle}s \texttt{ResNet-50} architecture out-performing the segmentation approach of U-Net in \texttt{Naberrie}. \texttt{MRC-3} performs similarly to the pre-trained \texttt{Zoobot2} network which is fine-tuned on a similar data set. It is worth noting that U-Net is a more complex architecture than the Mask R-CNN architecture, and thus the performance of \texttt{Naberrie} might improve with a larger training data set.

Figure\,\ref{fig:ROCAUC} (c) presents the classifier performance of the two \texttt{ResNet}-based architectures, \texttt{NG} and \texttt{ResNet-18} described in Sect.\,\ref{ssec: NG} , shown in red and blue, respectively, compared to the \texttt{4-layer CNN}, described in Sect.\,\ref{ssec:4layer}, shown in green. The performance of these three networks is relatively similar, yet better than a random classifier, with the \texttt{4-layer CNN} slightly out-performing the \texttt{ResNet}-based networks, with an AUROC score of $0.77$ as opposed to $0.71$ and $0.65$ for \texttt{NG} and \texttt{ResNet-18}, respectively. 

The networks with a commensurate classifier performance to that of \texttt{Zoobot1} are \texttt{Denselens}, and \texttt{LensCLR}, shown in Fig.\,\ref{fig:ROCAUC} (d), with AUROC scores of $0.83$ and $0.87$, shown in red and green, respectively. Like \texttt{Zoobot1}, these networks are more confident in identifying lens candidates and ranking those higher than its non-lens counterpart. \texttt{LensCLR} uses a semi-supervised approach, as opposed to the fully-supervised training of all other networks presented in this work. This has proven beneficial in the {network\textquotesingle}s performance since it has been trained with and without labelled data. Additionally, \texttt{Denselens} uses an ensemble of networks which leads to the benefit of the capability to rank-order candidate lenses to reduce the stochastic nature of trained parameters in CNNs. As a result, \texttt{Denselens} benefits from its sophisticated architecture, achieving a higher AUROC score than \texttt{NG}, mentioned above, which is trained on similar ground-based $\textit{r}$-band data from KiDS. However, with the addition of the U-Net, \texttt{U-Denselens} does not outperform its \texttt{Denselens} base architecture, with an AUROC score of $0.73$, similar to that of \texttt{NG}.

The networks described in Sect.\,\ref{ssec:Leuzzi}, however, perform no better than a random classifier, shown in Fig.\,\ref{fig:ROCAUC} (e). The three networks, \texttt{VGG}, \texttt{ResNext} and \texttt{GNet}, are unable to classify candidate lenses higher than the non-lenses in the common test set, with \texttt{ResNext} achieving the highest AUROC score of $0.54$, while \texttt{VGG} and \texttt{GNet} give AUROC scores of $0.51$ and $0.53$, respectively. We recapitulate that the training sample for a network has a significant impact in the {network\textquotesingle}s performance, yet the network architecture, training set size, and the complexity of the network can also play a crucial role.  A more in-depth analysis would need to be taken to investigate the performance of these networks

Finally, Fig.\,\ref{fig:ROCAUC} (f) shows the ROC curves for \texttt{CBEN} evaluated at two voting values $\beta$, namely \texttt{CBEN-0.3} and \texttt{CBEN-0.0001}, shown in red and blue, respectively. Although \texttt{CBEN-0.0001} performs better than its counterpart \texttt{CBEN-0.3}, its increased performance is not large, achieving an AUROC of $0.76$ in contrast to $0.72$ for \texttt{CBEN-0.3}. Although both \texttt{CBEN} networks perform better than other networks in this study, their overall performance is only at most $76\%$ confident at classifying a candidate lens higher than a non-lens. Though using a different training sample could increase the classifier performance of \texttt{CBEN}, despite the ensemble of seven CNNs assembling its sophisticated architecture, it is not as confident at classifying candidate lenses as other networks in this study of the Perseus cluster. 

Therefore, \texttt{Zoobot1}, \texttt{Denselens}, \texttt{4-layer CNN}, and \texttt{LensCLR} are the best overall classifiers for this application. 

\subsection{\text{TPR} and \text{FPR}}\label{ssec:tp}
The \text{TPR} and \text{FPR}, described in Sect.\,\ref{ssec:tpr fpr}, are of importance in lens finding, since the number of true positives is generally low due to the fact that strong lensing affects $\approx 1$ in \num{10000} galaxies. Meanwhile, the cost of a false positive is high. 
Ideally, the optimal network will have a high \text{TPR} and a low \text{FPR}, but often there is a trade-off between the two quantities, so we want a network that strikes a balance between the \text{TPR} and \text{FPR}, thus finding the majority, if not all, of the lens candidates in our common test set with few false positives. In this section, we compare the \text{TPR} and \text{FPR} of our best performing networks only, but the results for all networks are presented in Table\,\ref{table:1}.

\texttt{LensCLR} and \texttt{Zoobot1} have the highest AUC score of all networks, however, their performance is different with respect to \text{TPR} and \text{FPR}. Although \texttt{LensCLR} has the highest \text{TPR} of all networks finding $62/68$ lens candidates with a recall score of $94\%$, and recovers one more true positive than \texttt{Zoobot1}, it also finds nearly 5 times the number of false positives, leading \texttt{LensCLR} to have a much higher contamination ($N_\sfont{FP}$). Similarly, \texttt{MRC-95} and \texttt{4-layer CNN} also have high \text{TPR}s, finding $60/68$ and $48/68$ lens candidates, respectively, however, they also suffer from a large number of false positives.

\begin{figure}[t]%
    \centering
    {{\includegraphics[width=9cm]{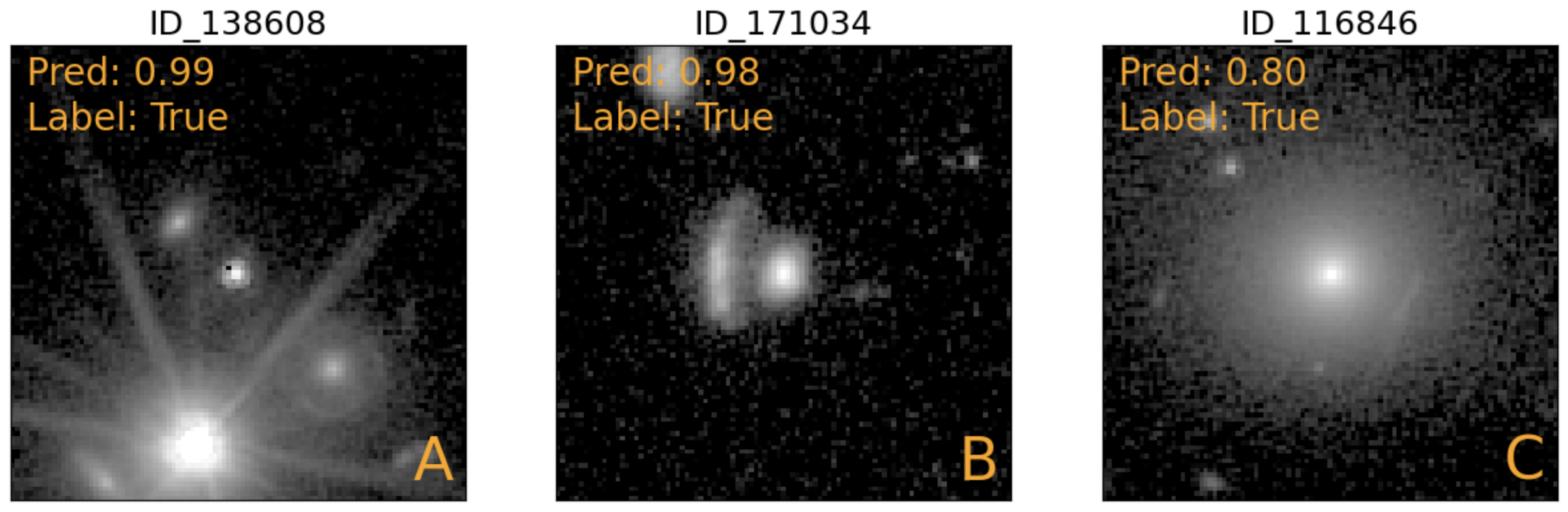} }}
    \caption{Examples of correctly identified candidate lenses (or true positives) by \texttt{Zoobot1} in the \IE-band. The grade of each candidate is shown at the bottom of each subplot. Each subplot contains the predicted score, $p_{\sfont{LENS}}$ for that candidate along with the true class (with True belonging to class  `lens'). Each candidate lens image is $10^{\prime\prime}\times10^{\prime\prime}$.}
    \label{fig:Zoobot TP}
\end{figure}

\begin{table*}
\caption{Performance metrics for the various models with their $p_{\sfont{THRESH}}$ values, based on the common test set of \num{12086} stamps. The true positive rate \text{TPR} and false positive rate \text{FPR} are given as percentages; total numbers of false positives ($N_\sfont{FP}$), true positives ($N_\sfont{TP}$), false negatives ($N_\sfont{FN}$), true negatives ($N\sfont{TN}$), and the contamination rate ($N_\sfont{FP}/N_\sfont{TP}$) are also shown. For a detailed description of each network, we refer the reader to the respective section listed in column 2. }
\centering


\begin{tabular}{ p{2.5cm} P{1.0cm} P{1.5cm} P{1.3cm} P{1.3cm} P{1.5cm} P{1.0cm} P{1.0cm} P{1.5cm} P{1.5cm}}

 Model& Section & $p_{\sfont{THRESH}}$ & TPR & FPR & $N_\sfont{FP}$ & $N_\sfont{TP}$ & $N_\sfont{FN}$ & $N_\sfont{TN}$ & $N_\sfont{FP}/N_\sfont{TP}$\\[1pt]
 \hline 
  & & & & & & & & \\[-8pt]
 \texttt{Zoobot1} & \ref{ssec:Zoobot} & 0.50 & 89.71 & \phantom{0}4.55 & \phantom{0}547 & 61 & \phantom{0}7 & 11471 & \phantom{0}\phantom{0}8.97\\
 \texttt{Zoobot2} & \ref{ssec:Zoobot} & 0.50 & 61.72 & \phantom{0}5.48 & \phantom{0}659 & 42 & 26 & 11359 & \phantom{0}15.69\\
 \texttt{E1} & \ref{ssec:Zoobot} & 0.50 & 41.84 & 10.09 & 1213 & 29 & 39 & 10805 & \phantom{0}41.83\\
 \texttt{E2} & \ref{ssec:Zoobot} & 0.50 & 45.42 & \phantom{0}9.09 & 1093 & 24 & 44 & 10925 & \phantom{0}45.54\\
 \texttt{Naberrie} & \ref{ssec:Wilde} & 0.90 & 26.47  & 23.64 & 2841 & 18 & 50 & \phantom{0}9177 & 157.83\\
 \texttt{MRC-95} & \ref{ssec:Wilde} & 0.95 & 80.24 & 65.14 & 7828 & 60 & \phantom{0}8 & \phantom{0}4190 & 130.47\\
 \texttt{MRC-99} & \ref{ssec:Wilde} & 0.95 & 48.53 & 20.51 & 2465 & 33 & 35 & \phantom{0}9553 & \phantom{0}74.70\\
 \texttt{MRC-995} & \ref{ssec:Wilde} & 0.95 & 60.29 & 38.74 & 4656 & 41 & 27 & \phantom{0}7362 & 113.56\\
 \texttt{MRC-3} & \ref{ssec:Wilde} & 0.95 & 48.53 & 20.51 & 2465 & 33 & 35 & \phantom{0}9553 & \phantom{0}74.70\\
 \texttt{Denselens} & \ref{ssec:Denselens} & 0.23 & 50.00 & \phantom{0}4.64 & \phantom{0}558 & 34  & 34 & 11460 & \phantom{0}16.41\\
 \texttt{U-Denselens} & \ref{ssec:Denselens} & 0.18 & 39.70 & \phantom{0}2.75 & \phantom{0}331 & 27  & 41 & 11687 & \phantom{0}12.26\\
 \texttt{NG} & \ref{ssec: NG} & 0.63& 63.00& 31.00& 3651& 43& 25& \phantom{0}8367 & \phantom{0}84.91\\
  \texttt{ResNet-18} & \ref{ssec: NG} & 0.99& 48.53& 22.36& 2703& 33& 35& \phantom{0}9315 & \phantom{0}81.91\\
 \texttt{4-layer CNN} & \ref{ssec:4layer} & 0.60 & 70.59 & 30.09 & 3616 & 48 & 20 & \phantom{0}8402 & \phantom{0}75.33\\  
 \texttt{LensCLR} & \ref{ssec: lensclr} & 0.50 & 94.00 & 21.00 & 2480 & 62 & \phantom{0}6 & \phantom{0}9515 & \phantom{0}40.00\\ 
\texttt{CBEN-0.3} & \ref{ssec:cben} & 0.30 & 66.18 & 22.42 & 2694 & 45 & 23 & \phantom{0}9324 & \phantom{0}59.87\\ 
\texttt{CBEN-0.0001} & \ref{ssec:cben} & 0.0001 & 39.71 & \phantom{0}4.85 & \phantom{0}584 & 27 & 41 & 11434 & \phantom{0}21.63\\ 
\texttt{VGG} & \ref{ssec:Leuzzi} & 0.30 & 20.60 & 21.29 & 2559 & 14 & 
 54 &  \phantom{0}9459 & 182.79\\ 
\texttt{GNet} & \ref{ssec:Leuzzi} & 0.30 & 22.06 & 21.98 & 2641 & 15 & 53 & \phantom{0}9377 & 176.07\\ 
\texttt{ResNext} & \ref{ssec:Leuzzi} & 0.30 & 29.41 & 22.70 & 2728 & 20 & 48 & \phantom{0}9290 & 136.40\\ 
\end{tabular}
\label{table:1}
\end{table*}
An example of correctly identified candidate lenses using \texttt{Zoobot1} for each grade, along with their predicted scores $p_{\sfont{LENS}}$ is shown in Fig.\,\ref{fig:Zoobot TP}. \texttt{Zoobot1} correctly identifies $61$ out of the $68$ candidate lenses giving a \text{TPR} or 
{recall} 
score of $\approx 90\%$. 
In Fig.\,\ref{fig:Zoobot TP}, $p_{\sfont{LENS}}$ is higher for candidate lenses graded $A$ and $B$ than for those in grade $C$. For a full distribution of $p_{\sfont{LENS}}$ across all grades $A$, $B$, and $C$ for \texttt{Zoobot1}, we refer the reader to Fig.\,\ref{fig:Zoobot Distribution} (left) which shows the grade $A$ and $B$ lens candidates at a higher predicted score than the $C$ grade candidates. This is in agreement with the results from the expert visual inspection where detailed lens models were fitted for all three grade $A$ lenses and two grade $B$ lenses. \texttt{Zoobot1} is confident in correctly identifying grade $A$ lenses and even the grade $B$ lenses, but less confident when correctly identifying grade $C$ lenses. In contrast, \texttt{Zoobot2} recovers only $42$ of the $68$ lens candidates giving a 
recall
score of only $62\%$ which is less than the recovered candidates of \texttt{Zoobot1}. 
For a detailed distribution of $p_{\sfont{LENS}}$ predictions from \texttt{Zoobot1} and \texttt{Zoobot2}, respectively, across grades $A$, $B$, and $C$, see 
Fig.\,\ref{fig:Zoobot Distribution}. 


We are particularly interested in the \text{FPR} and the contamination of the results ($N_\sfont{FP}$). For our common test set, only about two in a thousand objects are candidate lenses, meaning that the contamination will be high unless the \text{FPR} is much less than the \text{TPR}. 
From Table\,\ref{table:1}, \texttt{Zoobot1} has an \text{FPR} of $4.55\%$ and, in total, falsely identifies $547$ non-lenses as lens candidates.


Likewise to \texttt{Zoobot1}, \texttt{Denselens} and \texttt{CBEN-0.0001} have low \text{FPR}s of $4.64\%$ and $4.85\%$, and find $558$ and $584$ false positives, respectively. Although this is a similar performance to \texttt{Zoobot1}, \texttt{Denselens} has a balance between the number of true positives and false positives, giving a \text{TPR} of just $50\%$, which is considerably less. Similarly, \texttt{CBEN-0.0001} also has a low \text{TPR} of just $39.7\%$, again showing that reducing the number of false positives can also cause a decrease in recovering true positives. \texttt{U-Denselens}, however, with a lower $p_{\sfont{THRESH}}$ value of $0.18$ finds far fewer false positives than all other networks and achieves the lowest \text{FPR} of all networks, finding $331$ false positives in total. Regardless of a low number of false positives, \texttt{U-Denselens} has a low \text{TPR} and recalls only $39.7\%$ of the candidate lenses, showing that a good balance between true positives and false positives  is still required for a pure and complete lens sample. Since \texttt{Denselens} and \texttt{U-Denselens} have been trained from ground-based KiDS data without fine-tuning, unlike e.g. \texttt{Zoobot1}, we expect the performance of these networks to improve when trained on real \Euclid imaging in the future.

Overall, \texttt{U-Denselens} and \texttt{LensCLR} achieve the lowest FPR and highest recall, respectively. Although \texttt{U-Denselens} finds the least false positives, it recovers less than $50\%$ of the candidate lenses (true positives). In contrast to that, \texttt{LensCLR}, which recovers the highest number of candidate lenses of all networks, finds a large number of false positives with a high value of the FPR. Therefore, we are interested in a network with a good balance between completeness and purity, i.e. a network with a low FPR and high recall score. This condition is satisfied with the \texttt{Zoobot1} network, which has the second lowest FPR of all networks, and a high recall score, finding just one less candidate lens than \texttt{LensCLR}. Thus, \texttt{Zoobot1} is the best overall gravitational lens classifier in terms of a good balance between a low \text{FPR} at a high \text{TPR} in this application to the Perseus cluster imaging data. This is to be expected, since \texttt{Zoobot1} has a learned representation of galaxy morphology from over $100$\,\text{M} GZ classifications, in addition to simulated gravitational lenses that are painted onto real data from other \Euclid ERO fields. We can confidently state that fine-tuning a \texttt{Zoobot} model by adapting the head to solve a new task is currently the optimum method of lens finding for \Euclid. Additionally, we note that increasing the $p_{\sfont{THRESH}}$ value, rather than using the default $p_{\sfont{THRESH}}=0.5$, would further decrease the \text{FPR} for \texttt{Zoobot1}. Increasing the 
$p_{\sfont{THRESH}}$ value for this application, however, would also result in a reduced \text{TPR}, although it is possible that many grade $B$ and grade $C$ candidates are non-lenses, and lens modelling would be required to confirm this. 

\subsection{Common lens misclassifications}\label{ssec:fps}
Often, binary classifiers suffer from substantial false-positive classifications, or a high \text{FPR} in this field of study. In the context of lens identification, such false positives frequently occur due to images containing multiple sources, galaxies with arms or spiral galaxies, and also images with a central source and smaller surrounding sources. Therefore, a classifier might confuse these circumstances with lensing arcs or an Einstein ring.  

The common galaxy morphologies occurring as false positives in the classification by all networks are spirals with prominent arms, edge-on spirals, and stamps containing multiple sources. As stated in the previous section, \texttt{Zoobot1} finds a total of $547$ false positives. Examples of false positives occurring in {\texttt{Zoobot1}\textquotesingle}s classification are shown in 
Fig.\,\ref{fig:Zoobot1 FPs}. Figures\,\ref{fig:4-layer_CNN_FPs} and \ref{fig:U-Denselens_FPs} show example false positives found by \texttt{4-Layer CNN} and \texttt{U-Denselens}, respectively. It is a possibility that all networks are responding to the presence of prominent arcs and Einstein rings in candidate lenses, so spiral galaxies and edge-on spirals represent the perfect lens imposters. Prominent arms in spiral galaxies can closely approximate an arc or Einstein ring in lensed images, thus confusing the classifier. Additionally, it can be seen that samples containing multiple sources, e.g. multiple galaxies per stamp, and clumpy galaxies, are commonly occurring false positives. A possible explanation for such a misclassification is that multiple sources occasionally align themselves so as to form what the networks have learned to identify as gravitationally lensed source image morphologies.

To reduce the number of false positives, future studies could allow such commonly misclassified image types to constitute their own class, such that classifiers will be more easily able to learn the difference between images that contain real lenses, and images that contain lens imposters. The false positives from all networks are still recognisable as such by human experts, suggesting that machine learning improvements are still possible. An alternative possibility to simply improving the CNNs is to use a second, more interpretable or astrophysically-motivated stage, which could improve the purity of the lens candidates produced by CNNs. Nevertheless, \texttt{Zoobot1}, \texttt{Denselens} and \texttt{U-Denselens}, and \texttt{CBEN-0.0001}  achieve a good \text{FPR} for this first application to real data from the \Euclid ERO programme.

\subsection{Contamination rate}\label{ssec:Uniform Results}
In this section, we investigate the contamination rate, or the $N_\sfont{FP}/N_\sfont{TP}$ ratio, of all networks assuming a fixed \text{FPR} of $5\%$ for this application to the Perseus cluster data. The contamination rate is an important measure for estimating the number of sources that we would need to visually inspect for each network if applied to the remaining \Euclid ERO or future \Euclid observations. Although the contamination rate does not give the total number of candidates (i.e. $\text{TP}+\text{FP}$), it does provide the expected number of false positives a network will find per true positive. The choice of a low \text{FPR} of $5\%$ as a goal is practical, rather than e.g. $100\%$ completeness. Specifically, it recognises that a visual inspector has to check all candidates, and since there will be $\sim 1$ billion objects for the entire EWS, even a $5\%$ \text{FPR} results in a lot of objects to inspect. Once the threshold, $p_{\sfont{THRESH}}$, for each network to achieve this feasibly checkable number of candidates has been established, the networks can be compared on the basis of the completeness they achieve given those thresholds.

The \Euclid ERO programme covers 17 fields including the Perseus cluster and thus estimating the number of potential lens candidates from each network that would have to be inspected for the entirety of \Euclid ERO, and the cost of visual inspection time, is an important consideration for assessing and comparing the performance of the networks described in this work.  

CNNs are yet to be applied to the remaining $16$ \Euclid ERO fields which, in total, comprise a total data set size of $>\num{300000}$ sources. Applying CNNs with high contamination rates to the remaining ERO fields would result in a large number of false positives found, for which, in turn, a visual inspection of potential candidate lenses for verification becomes impractical. Here, we choose a $p_{\sfont{THRESH}}$ for each network which results in a \text{FPR} of $5\%$, and we compare the contamination rate for this application to our common test set of \num{12086} stamps from the Perseus cluster. We limit this comparison to the networks with a contamination rate, $N_\sfont{FP}/N_\sfont{TP}<100$, which 
leaves 
13 of our networks. A contamination rate $>100$ would require a visual inspection of $>$$100$ potential candidate lenses for every true lens, which is infeasible.

\begin{figure}[t]%
    \centering
    {{\includegraphics[width=9.1cm]{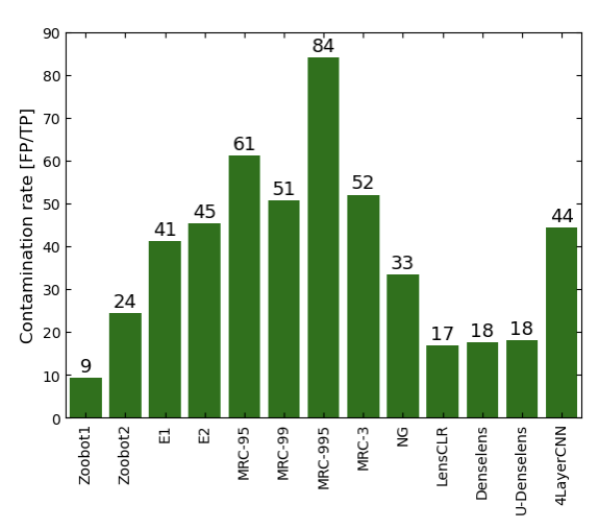} }}
    \caption{Contamination rate, or $N_\sfont{FP}/N_\sfont{TP}$ ratio, of the potential candidate lens sample of each network assuming a fixed \text{FPR} of $5\%$. Networks with a contamination rate $>100$ are not included in this comparison. The value given above each network shows the number of false positives found for every true candidate lens in the Perseus cluster data. }
    \label{fig:Contamination}
\end{figure}

Figure\,\ref{fig:Contamination} shows the contamination rate for each network in this work applied to the Perseus cluster data for a fixed \text{FPR} of $5\%$ with $N_\sfont{FP}/N_\sfont{TP}<100$. As expected, \texttt{Zoobot1} has the lowest contamination rate of all networks, finding nine false positives per true positive lens in the Perseus cluster. Additionally, \texttt{LensCLR}, \texttt{Denselens}, and \texttt{U-Denselens} find $17$ and $18$ false positives per true positive, respectively. Although this number of false positives per true positive might seem much higher than those found by \texttt{Zoobot1}, if we assume that an expert visual inspection of $\num{1000}$ potential lens candidates takes $1\text{h}$, then the time to verify the candidate lenses is $\approx 1\text{h} 10\text{min}$, which, for the entirety of the Perseus cluster, is feasible. Though with an increase in the time to visually inspect potential candidate lenses, \texttt{Zoobot2} and \texttt{NG} find $24$ and $33$ false positives, giving a total of $\num{1632}$ and $\num{2244}$ false positives for the Perseus cluster, respectively.

The remaining networks, however, have much higher contamination rates ranging from $\approx 40$ to $84$ false positives for every true positive. These high contamination rates come at the expense of an increased visual inspection time, which would take expert visual inspectors many hours to verify as non-lenses. In Sect.\,\ref{ssec:tp}, the \texttt{4-layer CNN} performed comparatively with \texttt{Denselens} network in terms of AUROC score, however, here it has a contamination rate of $>40$ false positives for every true positive. Although the number of false positives per true positive found are less than those found by other networks in this section, and the time to visually inspect the total false positives exceeds that of \texttt{Zoobot1}, \texttt{LensCLR}, and \texttt{Denselens}, inspecting roughly $40$ potential candidate lenses per true lens is reasonably feasible for the Perseus cluster. 

In summary, the networks presented in this work with the lowest contamination rates are \texttt{Zoobot1}, \texttt{LensCLR}, and \texttt{Denselens}, along with its additional U-Net \texttt{U-Denselens}, finding fewer false positives per one true positive than the remaining networks in this work.

\section{Discussion and conclusion}\label{sec: Conclusions}
We have presented a systematic comparison and first attempt of applying several networks, trained from different and diverse data sets, to search for strong gravitational lens candidates, tested on a common data set from a visual inspection of candidates in \Euclid ERO data of the Perseus cluster. The networks were trained on a variety of training data sets consisting of simulated lenses and simulated non-lenses, as well as real non-lenses. We used standard metrics to present the results such as ROC curves and AUROC score in Sect.\,\ref{ssec:rocauc}, and compared examples of false positives encountered by each network. We additionally made a comparison of the results to compare the contamination rate in Sect.\,\ref{ssec:Uniform Results}, or the $N_\sfont{FP}/N_\sfont{TP}$ ratio, to facilitate the number of potential candidate lenses from each network that would require expert visual inspection for verification.

The main conclusions are as follows.
\begin{itemize}
    \item Many networks trained from their own data sets, 
    comprising 
    simulated lenses and non-lenses and/or real lenses and non-lenses, suffer from poor performance when tested on real data, finding many false positives for which a visual inspection of candidates becomes infeasible.
    \item \texttt{Zoobot1} is the overall best performing network with a balance between a high \text{TPR} and low \text{FPR}, which has been pre-trained on $800$\,\text{k} galaxy images with over $100$\,\text{M} volunteer classifications and further fine-tuned on real data from other ERO fields.
    \item \texttt{LensCLR} achieves the highest recall score of all networks but this comes with the cost of a large number of false positives. Therefore, increasing the recall score can cause an increase in contamination.
    \item \texttt{Denselens} achieves a low \text{FPR} with \texttt{U-Denselens} achieving the lowest \text{FPR} of all networks, with both networks trained from ground-based data without fine-tuning, yet with low recall scores. It is evident that reducing the \text{FPR} can decrease the number of true positives recovered.
    \item \texttt{E1} and \texttt{E2}, with the same architecture as \texttt{Zoobot1}, under-perform on the common test set, implying that the nature of training, and the training sample, play a crucial role, and fine-tuning on a combination of real data and simulated gravitational lenses ultimately reduces the \text{FPR}. This requires further investigation.
    \item The common false positives shared by all networks are spiral galaxies with prominent arms, edge-on spirals, and multiple sources, which we refer to as lens imposters. 
\end{itemize}
We therefore recommend \texttt{Zoobot1}, \texttt{LensCLR}, \texttt{Denselens}, and \texttt{U-Denselens} to be used for future strong gravitational lens searches, with \texttt{Zoobot1} being the optimal network for reducing the FPR whilst maintaining a high true positive recovery, of which is of high importance in strong lens finding. However, 
the roughly $9:1$ 
false positive to true positive 
ratio in even our best CNN lens finding model implies that the CNNs are currently still far from being able to produce a pure sample of the $>\num{100000}$ strong gravitational lensing systems widely predicted for \Euclid. As previously stated, lensing affects $\approx 1$ in \num{10000} galaxies, so for the entire \Euclid ERO data, our best network \texttt{Zoobot1} with a $9:1$ false positive to true positive ratio would find $\approx 300$ lens candidates. Although a visual inspection of this number of candidates is easily attainable, we expect of the order of $>10^6$ candidates for the entirety of the EWS. This is a very large undertaking, but is still small in comparison to the size of e.g. the GZ data sets. Therefore, one approach to achieving a pure sample of $>10^5$ \Euclid lensing systems is to enrol the efforts of human (possibly expert) volunteers. The fact that the false positives are still quickly identifiable by experts, and the type of false positives found by all networks are in agreement, also suggests that improvements are possible in the machine learning lens candidate selection. Additionally, this work has focused on \IE-band data of the $\num{12086}$ stamps from the Perseus cluster only. The addition of the remaining \YE, \JE, and \HE filters offers further potential improvements to the machine learning lens candidate selection, which will be imperative in searching for strong gravitational lenses in \Euclid. 

\begin{acknowledgements}
\AckERO
R. Pearce-Casey thanks the Science and Technology Facilities Council (STFC) for support under grant ST/W006839/1.
V.B. and C.T. acknowledge the INAF grant 2022 LEMON.
A.M.G. acknowledges the support of project PID2022-141915NB-C22 funded by MCIU/AEI/10.13039/501100011033 and FEDER/UE.
M.W. is a Dunlap Fellow. The Dunlap Institute is funded through an endowment established by the David Dunlap family and the University of Toronto.
SHS thanks the Max Planck Society for support through the Max Planck Fellowship. This project has received funding from the European Research Council (ERC) under the European Union’s Horizon 2020 research and innovation programme (LENSNOVA: grant agreement No 771776). 
This research is supported in part by the Excellence Cluster ORIGINS which is funded by the Deutsche Forschungsgemeinschaft (DFG, German Research Foundation) under Germany’s Excellence Strategy -- EXC-2094 -- 390783311.
This work made use of \texttt{Astropy}: a community-developed core Python package and an ecosystem of tools and resources for astronomy \citep{astropy13,astropy18,astropy22}, \texttt{NumPy} \citep{numpy} and \texttt{Matplotlib} \citep{matplotlib}.


  \AckEC
\end{acknowledgements}

%
%

\bibliography{biblio, Euclid}

%
%

\appendix
  \onecolumn 
  
\section{Lens candidates}\label{apdx:A}

\begin{figure*}[!htb]%
    \centering
    {{\includegraphics[width=17.96cm]{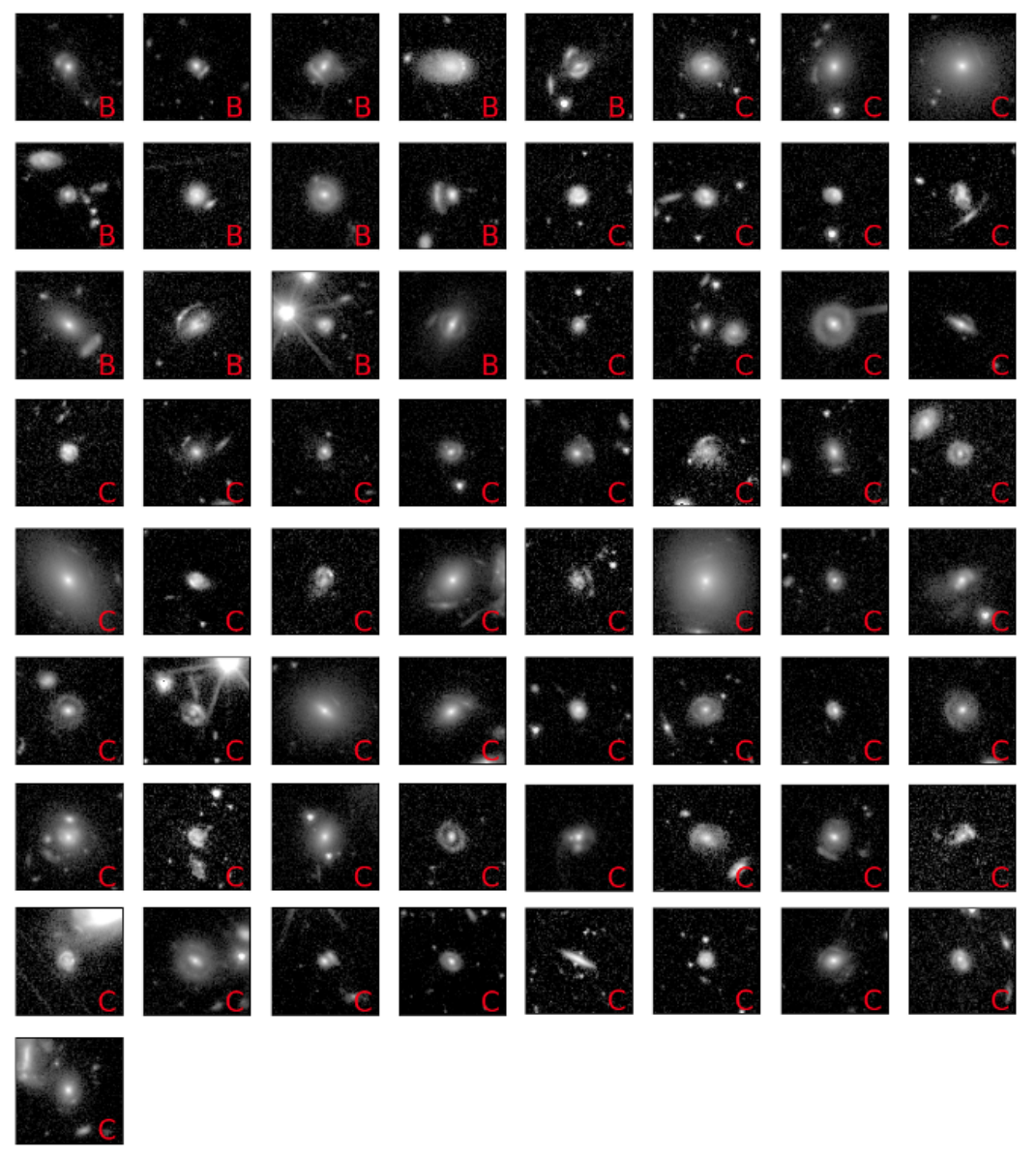}}}
    \caption{The $65$ potential lens candidates considered as true positives in the common test data set in this work. The three grade $A$ candidates are shown in Fig.\,\ref{fig:grade a} in the main text. $13$ grade $B$ and $52$ grade $C$ lens candidates are shown above with the respective grades shown in red in the bottom left corner of each subplot. Each lens candidate image is $10^{\prime\prime}\times10^{\prime\prime}$.}
    \label{fig:ABC}
\end{figure*}

We show the remaining $65$ combined grade $B$ and $C$ lens candidates out of the total $\num{12086}$ stamps with \IE$<23$ mag from the Perseus cluster ERO imaging in Fig.\,\ref{fig:ABC}. These candidates were visually inspected by professional astronomers and full details are given in \cite{barroso2024euclidearlyreleaseobservations}. There were $13$ candidates taken to be probable lenses graded $B$. Detailed lens models were fitted to two grade $B$ candidates which were consistent with a single source lensed by a plausible mass distribution. There were $52$ grade $C$ candidates, all of which have lensing features present but this could be explained by other physical phenomena. However, we have included all grade $C$ candidates in our common test set. 

\section{Distribution of predicted scores from fine-tuning}\label{apdx:B}

\begin{figure}[!htb]%
    \centering
    {{\includegraphics[width=18.2cm]{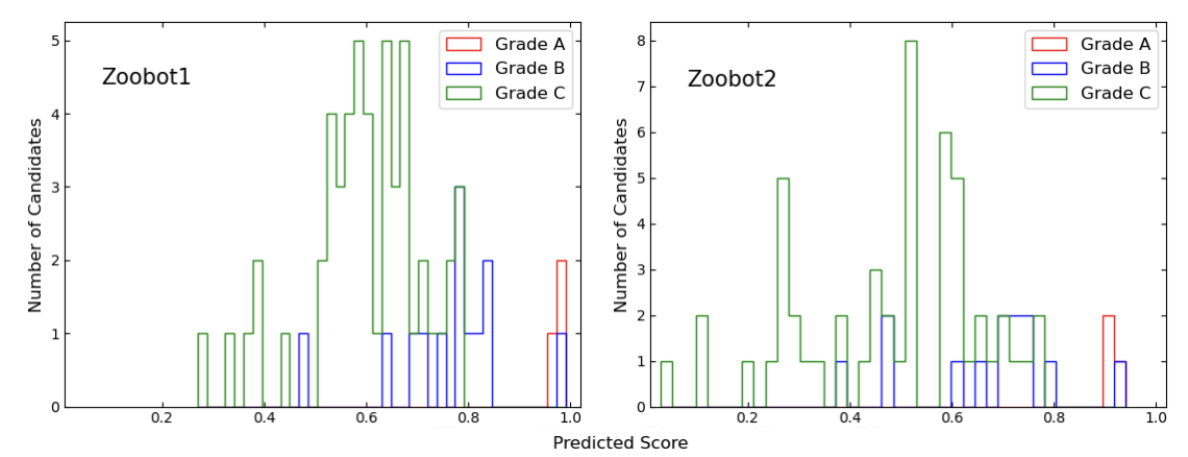} }}
    \caption{Distribution of predicted scores by \texttt{Zoobot1} (left) and \texttt{Zoobot2} (right) for lens candidates across grades $A$, $B$, and $C$. Grade $A$ lens candidates are confidently identified by both networks, shown to the right of the distribution in red at a high value of $p_{\sfont{LENS}}$. Grade $B$ predicted scores are shown in blue and are more varied for \texttt{Zoobot2}. The predicted scores for grade $C$ lens candidates have the most variation amongst grades, shown in green, for both networks illustrating that \texttt{Zoobot1} and \texttt{Zoobot2} are not as confident in identifying grade $C$ lens candidates.}
    \label{fig:Zoobot Distribution}
\end{figure}

We show the distribution of predictions, $p_{\sfont{LENS}}$, on the $68$ lens candidates given by the two fine-tuned \texttt{EfficientNetB0} networks with $p_{\sfont{THRESH}}=0.5$, namely \texttt{Zoobot1} and \texttt{Zoobot2}, described in Sect.\,\ref{ssec:Zoobot}. Figure\,\ref{fig:Zoobot Distribution} (left) shows the histogram of predicted scores for the $68$ lens candidates using \texttt{Zoobot1}, with the grade $A$ and $B$ candidates shown to the right of this distribution in red and blue, respectively. These candidates are confidently classified as lenses. There is one grade $B$ candidate that has a predicted score below the prediction threshold, thereby classified as a non-lens by \texttt{Zoobot1}. Figure\,\ref{fig:Zoobot Distribution} (right) shows the predicted scores for the lens candidates using \texttt{Zoobot2}. Although \texttt{Zoobot2} correctly classifies all grade $A$ definite strong lens candidates, it is not as confident in its prediction of the grade $B$ candidates. The predictions of the grade $C$ candidates vary the most for both networks, shown in green. This is in agreement with the fact that grade $C$ candidates are less likely to be lenses.

\section{False positives}\label{apxc}
\begin{figure}[!htb]%
    \centering
    {{\includegraphics[width=18.2cm]{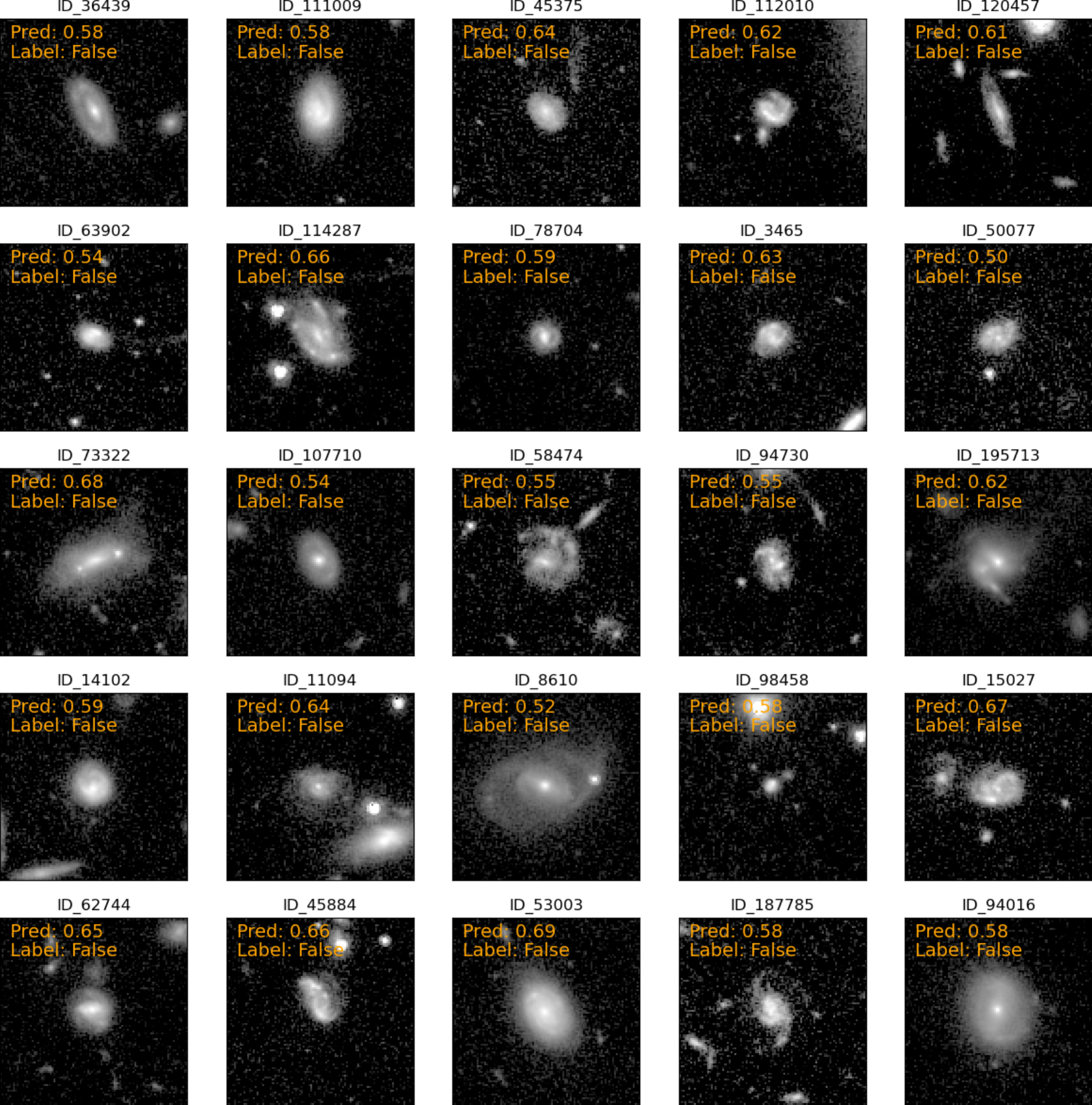} }}
    \caption{Example of $25$ out of $547$ false positives ($N_\sfont{FP}$ in Table \ref{table:1}) found by \texttt{Zoobot1} in the common test set. Each panel contains the predicted score, $p_{\sfont{LENS}}$, for each false positive, along with the true class (with False belonging to class `non-lens'). Common morphologies include spiral galaxies with prominent arms, edge-on spirals, and stamps containing multiple sources. Each \IE-band image is $10^{\prime\prime}\times10^{\prime\prime}$.}
    \label{fig:Zoobot1 FPs}
\end{figure}

\begin{figure}[!htb]%
    \centering
    {{\includegraphics[width=18.6cm]{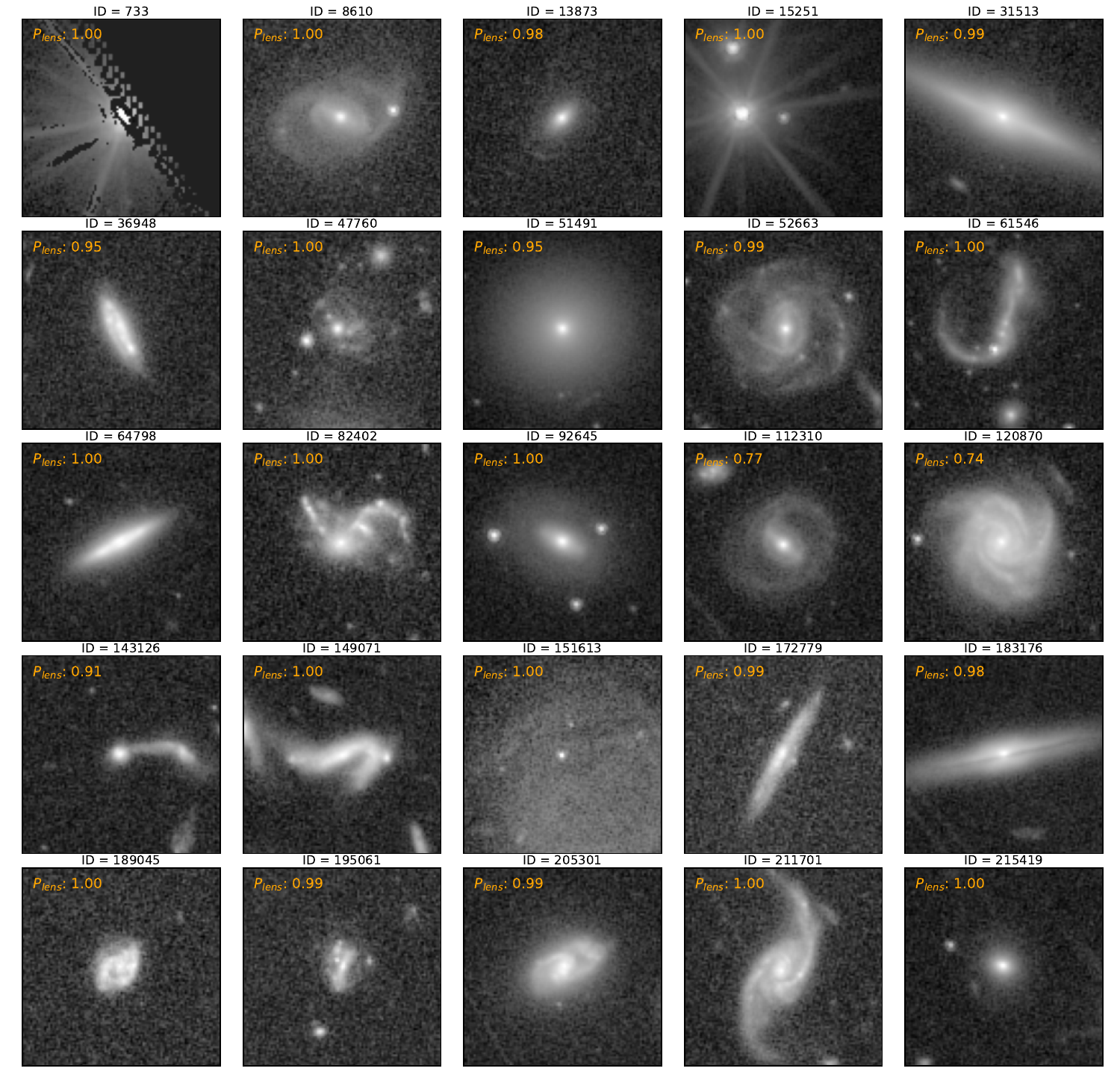} }}
    \caption{Example of $25$ out of the $3068$ false positives ($N_\sfont{FP}$ in Table \ref{table:1}) found by the \texttt{4-layer CNN} with $p_{\sfont{THRESH}}$ $\geq$ $0.7$ in the common test set. Each panel contains the ID of the source and its corresponding predicted score, $p_{\sfont{LENS}}$, for each false positive. The false positive sample contains a significant amount of edge-on and face-on spiral galaxies with prominent arms, stamps with multiple sources, and stamps with isolated sources. There is also a small fraction of stamps that fall near the borders of the Perseus footprint or with defective pixels. Each \IE-band image is $10^{\prime\prime}\times10^{\prime\prime}$.}
    \label{fig:4-layer_CNN_FPs}
\end{figure}

\begin{figure}[!htb]%
    \centering
    {{\includegraphics[width=18.6cm]{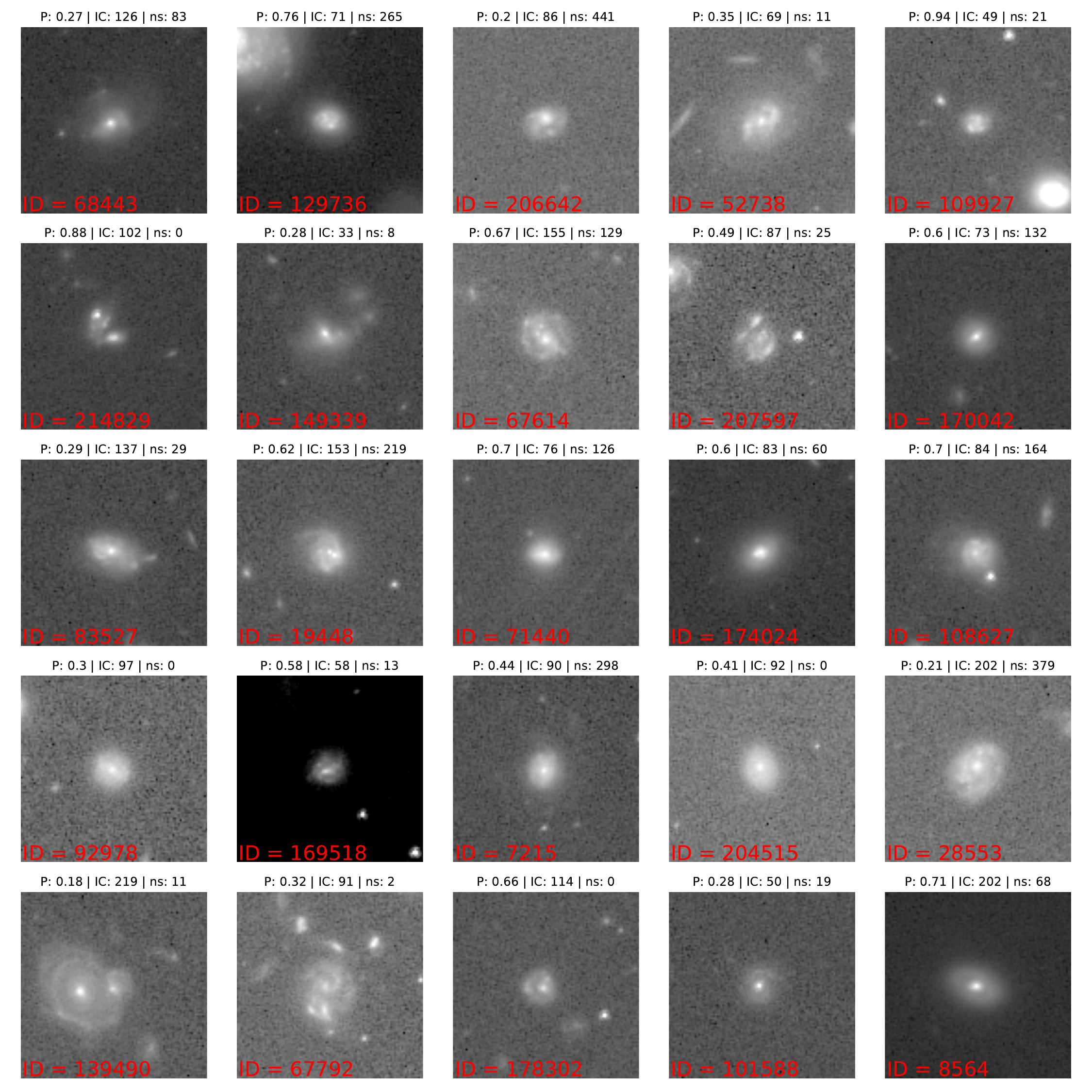} }}
    \caption{Example of $25$ out of the $331$ false positives ($N_\sfont{FP}$ in Table \ref{table:1}) found by the \texttt{U-Denselens} with the selection criteria in the common test set.  Each panel contains the predicted score, $p_{\sfont{LENS}}$, for each false positive along with the \text{IC} and $n_{\rm{s}}$. The ID of each source in shown in red in the bottom right. Frequently occurring morphologies are in agreement with other networks used in this work, including edge-on and face-on spiral galaxies, and stamps with multiple sources. Each \IE-band image is $10^{\prime\prime}\times10^{\prime\prime}$.}
    \label{fig:U-Denselens_FPs}
\end{figure}

Here, we show the common morphologies frequently occurring as false positives in the classification by all networks in this work. These morphologies include spiral galaxies, e.g. edge-on and face-on spirals, and spirals with prominent arms. Also common are stamps with multiple sources, and stamps with defective pixels that fall near the edges of the Perseus ERO footprint. Figure\,\ref{fig:Zoobot1 FPs} shows the false positives found by \texttt{Zoobot1} with the predicted score, $p_{\sfont{LENS}}$, shown in red in the top left corner of each subplot. Figure\,\ref{fig:4-layer_CNN_FPs} shows the false positives found by the \texttt{4-layer CNN} described in Sect.\,\ref{ssec:4layer}, which are common to the false positives found by \texttt{Zoobot1}, with the respective predicted scores shown in the top left corner of each subplot. Finally, Fig.\,\ref{fig:U-Denselens_FPs} shows the false positives found by \texttt{U-Denselens}, described in Sect.\,\ref{ssec:Denselens}, showing agreement in common morphologies that are misclassified as lenses to the other two networks shown here. 

\end{document}